\documentclass[10pt,twocolumn]{article}

\usepackage[T1]{fontenc}
\usepackage{lmodern}
\usepackage[margin=0.75in,top=0.85in,bottom=0.85in]{geometry}
\usepackage{amsmath,amssymb,amsfonts}
\usepackage{booktabs}
\usepackage{tabularx}
\usepackage{xcolor}
\usepackage{tikz}
\usetikzlibrary{shapes.geometric, arrows.meta, positioning, fit, calc, backgrounds}
\usepackage{listings}
\usepackage{url}
\usepackage[colorlinks=true,linkcolor=blue!60!black,citecolor=blue!60!black,urlcolor=blue!60!black]{hyperref}

\definecolor{aegisblue}{RGB}{28, 78, 128}
\definecolor{aegisgreen}{RGB}{34, 139, 34}
\definecolor{aegisred}{RGB}{180, 40, 40}
\definecolor{aegisgray}{RGB}{245, 247, 250}
\definecolor{codebg}{RGB}{248, 249, 252}

\title{\LARGE \textbf{Aegis OS: A Transactional Multi-Tenant Operating System Runtime for Eliminating Authorization Fatigue in Autonomous Coding Agents}}

\author{
  \textbf{Hui Zhao} \\
  Heurion Research \\
  \texttt{huizhao@heurion.org}
}

\date{Preprint --- October 2026}

\begin{document}

\maketitle

\begin{abstract}
Autonomous coding agents such as Google Antigravity (\texttt{agy}), Anthropic Claude Code (\texttt{claude}), and OpenAI Codex CLI (\texttt{codex}) execute multi-step software engineering workflows on developer workstations and cloud hosts. Existing agent execution environments face a \emph{Security--Autonomy Dilemma}: operators must choose between per-action confirmation prompts, which interrupt workflows on 45\%--57\% of tool calls in our benchmark trajectories and induce severe authorization fatigue, or unconfined execution (\texttt{--yolo} / \texttt{--dangerously-skip-permissions}), which exposes host files, SSH/cloud credentials, and API tokens to prompt injection and hallucination risks. Furthermore, concurrent agents editing a shared working tree clobber one another's changes, while passing large artifacts inline through LLM context windows incurs high token overhead.

We present \textbf{Aegis OS} (\texttt{aegis-os}), a transactional, multi-tenant Agent Operating System Runtime written in Rust and exposed via the Model Context Protocol (MCP) and the Google Agent2Agent (A2A) protocol. Aegis OS shifts the authorization boundary from individual tool calls to \emph{transactional workspace boundaries}. Each agent operates inside a kernel-enforced OS sandbox (macOS Seatbelt \texttt{sandbox-exec} and Linux Bubblewrap with default-deny \texttt{--tmpfs \$HOME} credential isolation, both evaluated, plus a microVM control-plane interface) backed by a Copy-on-Write (CoW) shadow workspace (APFS \texttt{libc::clonefile} on macOS; single-open \texttt{ioctl(FICLONE)} reflinks and optional \texttt{BTRFS\_IOC\_SNAP\_CREATE\_V2} subvolume snapshots on Linux), an identity-pinned HTTP Egress Proxy with L7 method/path-scoped boundary secret injection and outbound DLP payload inspection, and an A2A-compliant Inter-Agent Communication (IPC) bus paired with Content-Addressable Storage (CAS). Inside its isolated CoW workspace and organization IPC bus, an agent executes with zero confirmation prompts; human approval or TTL-scoped session leases are required only when committing diffs to the host via an LCS-based Diff3 3-Way Merge Engine and optional sandboxed pre-commit semantic verification gate (\texttt{verify\_cmd}). Across 1,500 tool calls in 50 controlled trajectories and \textbf{28,087 tool calls across 52 real coding-agent session logs}, Aegis OS reduces human confirmation prompts by \textbf{94.19\%--99.88\%} (controlled) and \textbf{90.7\%--97.9\%} (real sessions, lowering IPR from 53.31\% under per-mutation prompting and 28.54\% under prefix allowlists to \textbf{1.12\%}). Across 80 end-to-end adversarial attacks in eight categories (including unknown custom \texttt{\$HOME} credential paths, build-system commit poisoning, and same-domain L7 confused-deputy and DLP egress abuse), Aegis OS lowers the attack success rate from \textbf{100.0\% to 0.0\%} on both macOS and Linux; a nine-row single-component ablation shows that switching off any component re-opens the exact category it guards. On APFS it clones a 1,000-file (78.1 MB) workspace in \textbf{6.3 ms} (\textbf{16$\times$ faster} than \texttt{cp -cR}), and on Linux btrfs in \textbf{15.6 ms}, while preserving 100\% of disjoint multi-agent edits and catching 10/10 overlapping line conflicts and 10/10 disjoint-line semantic conflicts.
\end{abstract}

\section{Introduction}

The paradigm of AI-assisted software engineering has shifted from passive single-line code completion to autonomous, long-horizon coding agents---such as Google Antigravity (\texttt{agy}), Anthropic Claude Code (\texttt{claude}), OpenAI Codex CLI (\texttt{codex}), SWE-agent~\cite{yang2024sweagent}, and OpenHands~\cite{wang2025openhands}. Unlike conversational assistants, coding agents iteratively explore repositories, modify files across multiple modules, invoke compilers and test suites, query external APIs, and coordinate subtasks with peer agents~\cite{jimenez2024swebench,anthropic2024mcp}.

Despite advances in model reasoning, the execution substrate for local coding agents remains caught between usability and safety. When agent CLIs mutate the developer's live working tree directly, every file write (\texttt{fs\_write}, \texttt{fs\_edit}) and shell command (\texttt{proc\_exec}) carries immediate side effects on the host. Consequently, agent frameworks rely heavily on \emph{action-level human-in-the-loop (HITL) confirmation prompts}.

\subsection{The Security--Autonomy Dilemma}

In practice, action-level prompting and existing in-place CLI sandboxes exhibit three systemic limitations:

\begin{enumerate}
    \item \textbf{Severe Authorization Fatigue:} A 30-step bug-fix, refactoring, or security-audit trajectory typically invokes 13 to 18 mutating or state-changing operations. Prompting the developer on 45\%--57\% of tool invocations (as measured in \S\ref{sec:eval}) fragments autonomous workflow continuity to streaks of 4--7 steps and habituates users to reflexively approve prompts without inspecting arguments---a well-documented failure mode in usable security~\cite{felt2012android,akhawe2013alice}.
    \item \textbf{Risk of Unconfined ``YOLO'' Mode and In-Place Mutation:} To avoid constant interruptions, developers frequently enable bypass flags (\texttt{--dangerously-skip-permissions}, \texttt{--yolo}) or rely on lightweight in-place CLI sandboxes (e.g., Claude Code or Codex native Seatbelt/Landlock modes). In unconfined mode, an indirect prompt injection~\cite{greshake2023not,debenedetti2024agentdojo,zhan2024injecagent} can overwrite files outside the repository, read host credentials (\texttt{\textasciitilde/.ssh/id\_rsa}, \texttt{\textasciitilde/.aws/credentials}, \texttt{\textasciitilde/.kube/config}), or exfiltrate raw API keys (\texttt{\$GITHUB\_TOKEN}) from process environment variables. Even when native CLI sandboxes restrict writes to the project directory, agents still mutate the live host working tree in place without transactional rollback or boundary secret injection.
    \item \textbf{Multi-Agent Workspace Collisions and Context Bloat:} Developers increasingly run heterogeneous agents in parallel (e.g., \texttt{agy} for architecture, \texttt{claude} for backend logic, and \texttt{codex} for test generation)~\cite{wu2023autogen,hong2024metagpt}. When sharing a live working directory without manual \texttt{git worktree} management, concurrent edits overwrite one another's changes, while passing large specifications inline through JSON/text messages bloats LLM context windows.
\end{enumerate}

\begin{table*}[!t]
\centering
\caption{\textbf{Comparison of Agent Execution Substrates} across isolation mechanisms, authorization friction, workspace storage, and multi-agent coordination primitives.}
\label{tab:substrates}
\vspace{2pt}
\resizebox{\textwidth}{!}{%
\begin{tabular}{lccccccc}
\toprule
\textbf{Execution Substrate} & \textbf{Workspace Setup} & \textbf{$N$-Agent Disk Cost} & \textbf{Prompt Rate (IPR)} & \textbf{OS Kernel MAC} & \textbf{Boundary Secret Inject} & \textbf{Transactional 3-Way Merge} & \textbf{Native IPC + CAS} \\
\midrule
Host-Interactive (Unconfined CLI) & In-Place (0 ms) & Shared Tree & High (45\%--57\%) & $\times$ (Prompt Only) & $\times$ (Raw Env Vars) & $\times$ (In-Place Write) & $\times$ \\
Host-YOLO (\texttt{--skip-permissions}) & In-Place (0 ms) & Shared Tree & 0.0\% & $\times$ (Unconfined) & $\times$ (Raw Env Vars) & $\times$ (Silent Clobber) & $\times$ \\
Native CLI Sandbox (\texttt{claude}/\texttt{codex}) & In-Place ($\sim$6 ms) & Shared Tree & Low--Med & $\checkmark$ (Seatbelt/Landlock) & $\times$ (Env in Sandbox) & $\times$ (Shared Working Tree) & $\times$ \\
Container-per-Agent (Docker)~\cite{jimenez2024swebench,wang2025openhands} & Container Init & Per-Container Layer & Low & $\checkmark$ (Linux NS/cgroups) & $\times$ (Env in Container) & $\times$ (Isolated Volume) & $\times$ \\
AIOS (LLM Scheduler Kernel)~\cite{mei2024aios} & In-Place & Shared Tree & Medium & $\times$ (No OS MAC) & $\times$ (No Egress Proxy) & $\times$ (No CoW Merge) & Partial \\
\textbf{Aegis OS (Ours)} & \textbf{0.71--6.28 ms (APFS CoW)} & \textbf{Modified Files Only} & \textbf{0.07\%--3.33\%} & $\mathbf{\checkmark}$ (\texttt{sandbox-exec}/\texttt{bwrap}) & $\mathbf{\checkmark}$ (\texttt{\$AOS\_SECRET\_*}) & $\mathbf{\checkmark}$ (2-Phase LCS Diff3) & $\mathbf{\checkmark}$ (\texttt{aos-ipc}) \\
\bottomrule
\end{tabular}%
}
\end{table*}

\subsection{Key Insight: Transaction-Boundary Security}

Inspired by operating system transactions~\cite{porter2009txos,nightingale2005speculator} and layered copy-on-write filesystem sandboxes~\cite{kim2013mbox,wright2004unionfs}, our key insight is that \emph{authorization prompts should not guard speculative intermediate actions inside an isolated state space; they should guard irreversible boundary crossings into the host repository or untrusted external networks}.

In Aegis OS, each agent task executes as a three-step workspace transaction (Table~\ref{tab:substrates}):
\begin{enumerate}
    \item \textbf{Step 1 --- Instant CoW Shadow Fork (0.7--6.3 ms on APFS):} When an agent session starts, Aegis OS clones the host repository into an isolated Copy-on-Write (CoW) shadow workspace via APFS \texttt{libc::clonefile}, sharing unchanged disk blocks with zero data copying.
    \item \textbf{Step 2 --- Green-Zone Speculation (0 Prompts):} Inside its shadow workspace, the agent freely edits files, runs compilers and tests inside a kernel OS sandbox, coordinates with peer agents over A2A/IPC, and queries allowlisted APIs using opaque secret placeholders (\texttt{\$AOS\_SECRET\_*}). Because the live host tree cannot be modified and plaintext credentials never enter the sandbox, Aegis OS auto-allows 100\% of in-sandbox operations with \textbf{zero confirmation prompts}.
    \item \textbf{Step 3 --- Boundary Commit Gate (1 Prompt or Lease):} Only when the agent finishes testing and invokes \texttt{workspace\_commit} does Aegis OS validate a two-phase 3-way merge against the live host repository and request human approval (or consume a time-bounded session lease).
\end{enumerate}

\subsection{Contributions}

We design, implement, and evaluate \textbf{Aegis OS} (\texttt{aegis-os}), a multi-tenant Agent Operating System Runtime in Rust:

\begin{itemize}
    \item \textbf{Transactional CoW VFS with Two-Phase LCS Diff3 Merge and Pre-Commit Verification (\S\ref{sec:vfs}):} \texttt{aos-vfs} uses APFS \texttt{libc::clonefile} syscalls to provision isolated shadow workspaces in metadata time (0.71--6.28 ms for 100--1,000 files on APFS; single-open \texttt{FICLONE} reflinks in 2.9--15.6 ms on Linux btrfs), supports millisecond snapshots and rollbacks while preserving compiler build caches (\texttt{CARGO\_TARGET\_DIR}) outside the rollback tree, and combines a two-phase Longest Common Subsequence (LCS) Diff3 3-Way Merge Engine~\cite{khanna2007formal} with an optional sandboxed pre-commit semantic verification gate (\texttt{verify\_cmd}) that catches both overlapping line conflicts and disjoint-line semantic regressions before any host file is modified.
    \item \textbf{Kernel Sandbox and L7 Capability-Scoped Egress Proxy (\S\ref{sec:sandbox}):} \texttt{aos-sandbox} and \texttt{aos-egress} enforce mandatory access control via macOS Seatbelt (\texttt{/usr/bin/sandbox-exec} with file-write, credential-read, AppleEvent, and exact proxy-port network confinement) and a Linux Bubblewrap (\texttt{bwrap}) namespace driver with a token-authenticated Unix-socket egress bridge, paired with a TCP-pinned HTTP Egress Proxy that injects \texttt{\$AOS\_SECRET\_*} tokens only when both the target domain and the canonicalized HTTP method + URL path prefix match the secret's capability scope (\texttt{SecretScopeRule}).
    \item \textbf{A2A-Compliant Inter-Agent IPC Bus and Reference-Based CAS (\S\ref{sec:ipc}):} \texttt{aos-ipc} aligns inter-agent coordination with the Google Agent2Agent (A2A) protocol~\cite{google2025a2a}, providing global peer discovery with kernel PID liveness pruning, dynamic \texttt{AgentCard} synthesis (\texttt{/.well-known/agent-card.json}), stateful A2A \texttt{Task} lifecycle tracking (\texttt{POST /a2a}), P2P mailboxes, topic Pub/Sub, and Content-Addressable Storage (\texttt{artifact://<org>/sha256:<hash>}) that reduces in-transit IPC payload tokens by 99.85\%.
    \item \textbf{Cedar-Inspired Policy Engine with Commit-Time Poisoning Guards (\S\ref{sec:policy}):} \texttt{aos-policy} implements a five-stage authorization hierarchy inspired by Cedar~\cite{cutler2024cedar}, combining host red-line protection and commit-time sensitive path guards (\texttt{.git/hooks}, \texttt{.envrc}, \texttt{.github/workflows}, \texttt{build.rs}, \texttt{Makefile}, \texttt{package.json}, \texttt{.cargo/config.toml}) with one-shot and TTL-scoped session leases.
    \item \textbf{Empirical and Real-Trace Evaluation (\S\ref{sec:eval}):} We evaluate Aegis OS on macOS and, in a virtual machine, on Linux (btrfs, XFS, ext4) across 1,500 controlled tool calls, \textbf{28,087 real tool calls from 52 coding-agent session transcripts}, 80 end-to-end red-team attacks in eight categories, a single-component ablation, multi-scale CoW provisioning compared against \texttt{cp}, \texttt{rsync}, and \texttt{git}, and multi-agent Rust compilations (\texttt{cargo test} and \texttt{cargo check}).
\end{itemize}

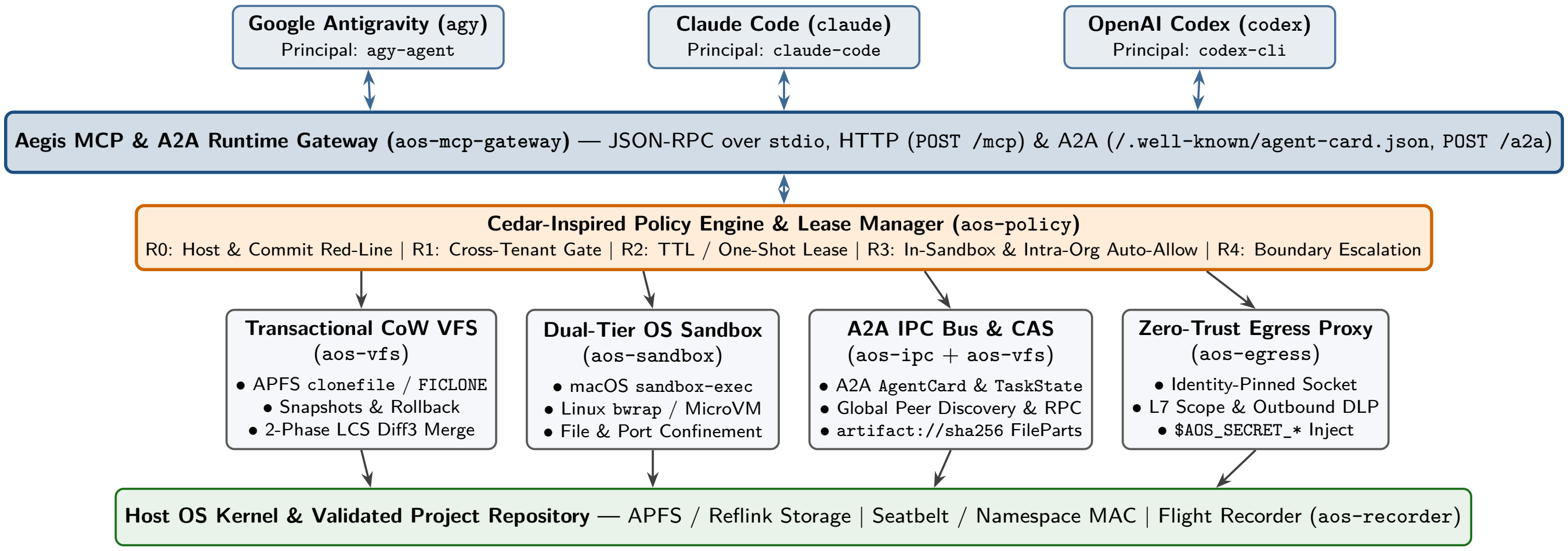
\begin{figure*}[!t]
\centering
\resizebox{\textwidth}{!}{%
\begin{tikzpicture}[
    font=\sffamily\footnotesize,
    node distance=0.45cm and 0.6cm,
    client/.style={rectangle, rounded corners=3pt, draw=aegisblue!80, fill=aegisblue!10, thick, minimum width=3.6cm, minimum height=0.75cm, align=center},
    gateway/.style={rectangle, rounded corners=3pt, draw=aegisblue, fill=aegisblue!20, very thick, minimum width=14.6cm, minimum height=0.8cm, align=center},
    policy/.style={rectangle, rounded corners=3pt, draw=orange!80!black, fill=orange!15, very thick, minimum width=14.6cm, minimum height=0.85cm, align=center},
    subsys/.style={rectangle, rounded corners=3pt, draw=black!65, fill=aegisgray, thick, minimum width=3.35cm, minimum height=1.85cm, align=center},
    host/.style={rectangle, rounded corners=3pt, draw=aegisgreen!80!black, fill=aegisgreen!10, thick, minimum width=14.6cm, minimum height=0.75cm, align=center},
    arrow/.style={-{Stealth[scale=0.9]}, thick, draw=black!75},
    biarrow/.style={{Stealth[scale=0.9]}-{Stealth[scale=0.9]}, thick, draw=aegisblue!85}
]

\node[client] (agy) {\textbf{Google Antigravity (\texttt{agy})}\\{\scriptsize Principal: \texttt{agy-agent}}};
\node[client, right=1.9cm of agy] (claude) {\textbf{Claude Code (\texttt{claude})}\\{\scriptsize Principal: \texttt{claude-code}}};
\node[client, right=1.9cm of claude] (codex) {\textbf{OpenAI Codex (\texttt{codex})}\\{\scriptsize Principal: \texttt{codex-cli}}};

\node[gateway, below=0.55cm of claude] (mcp) {\textbf{Aegis MCP \& A2A Runtime Gateway (\texttt{aos-mcp-gateway})} --- JSON-RPC over \texttt{stdio}, HTTP (\texttt{POST /mcp}) \& A2A (\texttt{/.well-known/agent-card.json}, \texttt{POST /a2a})};

\node[policy, below=0.4cm of mcp] (pol) {\textbf{Cedar-Inspired Policy Engine \& Lease Manager (\texttt{aos-policy})}\\{\scriptsize R0: Host \& Commit Red-Line $\mid$ R1: Cross-Tenant Gate $\mid$ R2: TTL / One-Shot Lease $\mid$ R3: In-Sandbox \& Intra-Org Auto-Allow $\mid$ R4: Boundary Escalation}};

\node[subsys, below=0.5cm of pol, xshift=-5.62cm] (vfs) {\textbf{Transactional CoW VFS}\\(\texttt{aos-vfs})\\[2pt]{\scriptsize $\bullet$ APFS \texttt{clonefile} / \texttt{FICLONE}}\\[-1pt]{\scriptsize $\bullet$ Snapshots \& Rollback}\\[-1pt]{\scriptsize $\bullet$ 2-Phase LCS Diff3 Merge}};

\node[subsys, right=0.38cm of vfs] (sbx) {\textbf{Dual-Tier OS Sandbox}\\(\texttt{aos-sandbox})\\[2pt]{\scriptsize $\bullet$ macOS \texttt{sandbox-exec}}\\[-1pt]{\scriptsize $\bullet$ Linux \texttt{bwrap} / MicroVM}\\[-1pt]{\scriptsize $\bullet$ File \& Port Confinement}};

\node[subsys, right=0.38cm of sbx] (ipc) {\textbf{A2A IPC Bus \& CAS}\\(\texttt{aos-ipc} + \texttt{aos-vfs})\\[2pt]{\scriptsize $\bullet$ A2A \texttt{AgentCard} \& \texttt{TaskState}}\\[-1pt]{\scriptsize $\bullet$ Global Peer Discovery \& RPC}\\[-1pt]{\scriptsize $\bullet$ \texttt{artifact://sha256} FileParts}};

\node[subsys, right=0.38cm of ipc] (egr) {\textbf{Zero-Trust Egress Proxy}\\(\texttt{aos-egress})\\[2pt]{\scriptsize $\bullet$ Identity-Pinned Socket}\\[-1pt]{\scriptsize $\bullet$ L7 Scope \& Outbound DLP}\\[-1pt]{\scriptsize $\bullet$ \texttt{\$AOS\_SECRET\_*} Inject}};

\node[host, below=0.5cm of sbx, xshift=1.87cm] (hostos) {\textbf{Host OS Kernel \& Validated Project Repository} --- APFS / Reflink Storage $\mid$ Seatbelt / Namespace MAC $\mid$ Flight Recorder (\texttt{aos-recorder})};

\draw[biarrow] (agy.south) -- ([xshift=-5.5cm]mcp.north);
\draw[biarrow] (claude.south) -- (mcp.north);
\draw[biarrow] (codex.south) -- ([xshift=5.5cm]mcp.north);

\draw[biarrow] (mcp.south) -- (pol.north);

\draw[arrow] ([xshift=-5.62cm]pol.south) -- (vfs.north);
\draw[arrow] ([xshift=-1.87cm]pol.south) -- (sbx.north);
\draw[arrow] ([xshift=1.87cm]pol.south) -- (ipc.north);
\draw[arrow] ([xshift=5.62cm]pol.south) -- (egr.north);

\draw[arrow] (vfs.south) -- ([xshift=-5.62cm]hostos.north);
\draw[arrow] (sbx.south) -- ([xshift=-1.87cm]hostos.north);
\draw[arrow] (ipc.south) -- ([xshift=1.87cm]hostos.north);
\draw[arrow] (egr.south) -- ([xshift=5.62cm]hostos.north);

\end{tikzpicture}%
}
\caption{Architecture of the Aegis OS Runtime (\texttt{aegis-os}). Coding agents connect via MCP (\texttt{stdio} or HTTP daemon) as distinct principals within a tenant organization. The policy engine auto-allows in-sandbox CoW mutations, confined process executions, and intra-organization IPC/CAS operations with zero prompts, while guarding host commits and network egress.}
\label{fig:arch}
\end{figure*}

\section{Threat Model and Evaluation Metrics}
\label{sec:threat}

\subsection{System and Adversary Model}

We consider a multi-agent deployment where autonomous coding agents operate within one or more tenant organizations. Each agent session is identified by its organization tenant (e.g., \texttt{aegis-os}), unique agent principal ID (e.g., \texttt{agy-agent}, \texttt{claude-code}, \texttt{codex-cli}), and role.

\paragraph{Trusted Computing Base and Enforcement Assumption.} We trust the host operating system kernel (macOS XNU/Seatbelt or Linux kernel namespaces) and the Aegis OS Runtime binary (\texttt{aegis-os}), which executes outside the agent sandbox. Because MCP is an additive tool protocol, our security guarantees assume that the host agent client is configured to route filesystem mutations and shell execution through \texttt{aegis-os} (i.e., built-in unconfined host tools such as raw \texttt{Bash} or host \texttt{Write} are either disabled in client permissions or intercepted by client \texttt{PreToolUse} hooks, as detailed in \S\ref{sec:integration}).

\paragraph{Adversarial Threat Vectors (T1--T5).} We assume the LLM driving any agent is untrusted and may issue malicious or erroneous tool calls due to:
\begin{itemize}
    \item \textbf{T1 (Indirect Prompt Injection):} Malicious instructions embedded in repository files, commit logs, issue trackers, or web pages read by the agent~\cite{greshake2023not,debenedetti2024agentdojo}.
    \item \textbf{T2 (Host Path Escape and Tampering):} Attempting to write files outside the workspace (e.g., \texttt{../../etc/hosts}, \texttt{\textasciitilde/.zshrc}, or sibling repositories) or staging commit-time hooks and build scripts (\texttt{.git/hooks/pre-commit}, \texttt{.envrc}, \texttt{build.rs}, \texttt{Makefile}).
    \item \textbf{T3 (Credential Harvesting):} Attempting to read sensitive host credentials (\texttt{\textasciitilde/.ssh/id\_rsa}, \texttt{\textasciitilde/.aws/credentials}, \texttt{\textasciitilde/.gnupg}, \texttt{\textasciitilde/.kube/config}, \texttt{\textasciitilde/.npmrc}, \texttt{\textasciitilde/.config/gcloud}, custom vault directories in \texttt{\$HOME}, or macOS Keychains).
    \item \textbf{T4 (Network Secret Exfiltration):} Attempting to transmit API tokens or source code to unauthorized external domains (\texttt{evil-exfil.org}, \texttt{webhook.site}) via HTTP requests or direct socket bypass (\texttt{curl --noproxy '*'}), sending a secret toward an allowlisted domain it is not bound to, or abusing disallowed L7 paths or request bodies on a bound domain.
    \item \textbf{T5 (Cross-Tenant Privilege Escalation):} Attempting to read CAS artifacts or inject IPC messages into a different tenant organization.
\end{itemize}

\paragraph{Non-Adversarial Concurrency and Reliability Hazards (H1--H2).} Even in benign settings, multi-agent workflows face: \textbf{(H1) Concurrent Lost Updates}, where parallel agents overwrite one another's edits on shared files; and \textbf{(H2) Multi-File Hallucinated Corruption}, where an agent breaks syntax across dozens of modules during speculative refactoring.

\subsection{Evaluation Metrics}

We evaluate authorization friction and security using four direct operational metrics:
\begin{itemize}
    \item \textbf{Interruption Prompt Rate (IPR):} The percentage of runtime tool calls that halt execution to request interactive human confirmation ($\text{Human Prompts} / \text{Total Tool Calls} \times 100\%$).
    \item \textbf{Modeled Human Verification Wait Time ($T_{\text{human}}$):} The cumulative human review overhead across a workload, assuming a conservative average context-switch and inspection time of $5.0\text{ s}$ per confirmation prompt.
    \item \textbf{Maximum Autonomous Streak (MAS):} The longest contiguous sequence of tool calls executed without a human confirmation prompt.
    \item \textbf{Attack Success Rate (ASR):} The percentage of adversarial attacks that successfully modify host files outside the workspace, read restricted credentials, exfiltrate secrets, or breach tenant isolation.
\end{itemize}

\section{Aegis OS Runtime Architecture}
\label{sec:arch}

Figure~\ref{fig:arch} illustrates the architecture of Aegis OS, and Table~\ref{tab:syscalls} summarizes its 35 MCP tools, 7 dynamic URI resources, and native Google A2A HTTP endpoints. Aegis OS follows an \emph{``Environment Runtime Only''} design: it does not embed LLM planners or prompt templates, instead providing virtual filesystem transactions, OS process isolation, network boundary proxying, A2A-compliant inter-agent coordination, and policy enforcement.

\begin{table*}[!t]
\centering
\caption{\textbf{Aegis OS MCP Tool, Dynamic URI Resource, and Google A2A Interface} (35 MCP Tools and 7 Dynamic URI Resources implemented in \texttt{aos-mcp-gateway}).}
\label{tab:syscalls}
\vspace{2pt}
\resizebox{\textwidth}{!}{%
\begin{tabular}{lll}
\toprule
\textbf{Subsystem} & \textbf{MCP Tools / Dynamic URI Resources / A2A Endpoints} & \textbf{Default Policy Gate \& Runtime Semantics} \\
\midrule
\textbf{Transactional VFS} (\texttt{aos-vfs}, 6 tools) & \texttt{fs\_read}, \texttt{fs\_write}, \texttt{fs\_edit}, \texttt{fs\_delete}, \texttt{fs\_glob}, \texttt{workspace\_status} & \textbf{Auto-Allow (0 Prompts)}: Confined to agent's CoW shadow tree \\
\textbf{Time-Travel, Sync \& Merge} (\texttt{aos-vfs}, 6 tools) & \texttt{workspace\_snapshot}, \texttt{workspace\_diff}, \texttt{workspace\_rollback}, \texttt{workspace\_sync}, \texttt{workspace\_commit}, \texttt{workspace\_export\_pr} & Snapshots/Rollback/Rebase-Sync \textbf{Auto-Allow}; \texttt{workspace\_commit} \& \texttt{workspace\_export\_pr} \textbf{Approval/Lease Gate} \\
\textbf{OS Process Sandbox} (\texttt{aos-sandbox}, 1 tool) & \texttt{proc\_exec} (macOS Seatbelt \texttt{sandbox-exec} / Linux \texttt{bwrap} / MicroVM spec) & \textbf{Auto-Allow (0 Prompts)}: Writes restricted to CoW root; blocks credentials \\
\textbf{Inter-Agent IPC \& Claims} (\texttt{aos-ipc}, 10 tools) & \texttt{ipc\_send}, \texttt{ipc\_recv}, \texttt{ipc\_request}, \texttt{ipc\_reply}, \texttt{ipc\_ack}, \texttt{ipc\_publish}, \texttt{ipc\_subscribe}, \texttt{ipc\_list\_peers}, \texttt{coord\_claim}, \texttt{coord\_release} & \textbf{Auto-Allow Intra-Org \& Workstation IPC}: Verified global peers (\texttt{<org>/<agent>}) allowed; file claims enforced \\
\textbf{Google A2A Protocol} (\texttt{aos-ipc}, 4 tools + HTTP) & \texttt{ipc\_update\_card}, \texttt{ipc\_task\_update}, \texttt{ipc\_task\_get}, \texttt{ipc\_task\_list} + \texttt{GET /.well-known/agent-card.json}, \texttt{POST /a2a} & \textbf{Auto-Allow Discovery \& Task RPC}: \texttt{AgentCard} synthesis \& stateful \texttt{A2ATaskState} lifecycle (\texttt{submitted}$\to$\texttt{completed}) \\
\textbf{Org CAS \& KV} (\texttt{aos-vfs}, 4 tools) & \texttt{org\_artifact\_publish}, \texttt{org\_artifact\_fetch}, \texttt{org\_kv\_put}, \texttt{org\_kv\_get} & \textbf{Auto-Allow Intra-Org}: Cross-org \texttt{artifact://} \texttt{FilePart} fetch requires policy boundary approval (\textbf{Ask}/\textbf{Deny}) \\
\textbf{Zero-Trust Egress} (\texttt{aos-egress}, 1 tool) & \texttt{egress\_http\_request} + Identity-Pinned TCP Proxy (\texttt{127.0.0.1:<port>}) & \textbf{Allowlisted Domains + L7 Scope Auto-Allow}: Unlisted domains escalate; injects \texttt{\$AOS\_SECRET\_*} \& runs DLP \\
\textbf{Policy \& Leases} (\texttt{aos-policy}, 3 tools) & \texttt{policy\_list\_escalations}, \texttt{policy\_approve\_escalation}, \texttt{policy\_grant\_lease} & Operator control plane for one-shot (\texttt{ttl=0}) or TTL-scoped \texttt{LeaseGrant} tokens \\
\textbf{Dynamic MCP URIs} (7 resources) & \texttt{aegis://workspace/status}, \texttt{aegis://ipc/inbox}, \texttt{aegis://org/peers}, \texttt{aegis://global/peers}, \texttt{aegis://a2a/agent-cards}, \texttt{aegis://a2a/tasks}, \texttt{aegis://org/shared/catalog} & Live state introspection for IDE status bars, global peer topology, and A2A task tables \\
\bottomrule
\end{tabular}%
}
\end{table*}

\subsection{Transactional CoW Virtual Filesystem and Two-Phase LCS Diff3 Merge (\texttt{aos-vfs})}
\label{sec:vfs}

To eliminate confirmation prompts on file mutations without risking host corruption, \texttt{aos-vfs} isolates every agent principal inside an independent Copy-on-Write shadow workspace rooted at:
\begin{center}
{\footnotesize\texttt{<runtime\_dir>/orgs/<org>/agents/<agent>/cow\_workspace/\allowbreak upper\_workspace/}}
\end{center}
(where \texttt{<runtime\_dir>} defaults to \texttt{/tmp/\allowbreak aegis-os-runtime} or \texttt{\textasciitilde/.aegis-os}), alongside a \texttt{baseline\_snapshot/} directory recording the host repository state at session initialization or last commit, and a \texttt{snapshots/<snap\_id>/} directory storing time-travel checkpoints.

\paragraph{APFS \texttt{clonefile} Shadow Provisioning.} Copying a multi-megabyte repository per agent via user-space read/write loops scales linearly with total byte volume. On macOS (APFS), \texttt{aos-vfs} invokes the native XNU syscall \texttt{libc::clonefile(src, dst, 0)} on eligible directory subtrees and files (automatically skipping heavy build/VCS directories such as \texttt{.git}, \texttt{target}, and \texttt{node\_modules}). Because \texttt{clonefile} duplicates only filesystem metadata extents and shares physical blocks until mutated, workspace initialization on APFS depends only on file count (0.71--6.28 ms for 100--1,000 files) and allocates physical storage only for modified files. Unlike \texttt{git worktree}, which requires a valid git commit history, shares the repository object store, and cannot snapshot dirty untracked trees in sub-millisecond time, \texttt{aos-vfs} operates transparently on any directory and supports instant nested checkpoints.

\paragraph{Linux Provisioning, Single-Open Syscall Reduction, and Btrfs Subvolume Snapshots.} Linux has no general directory-level clone: \texttt{ioctl(FICLONE)} reflinks one file at a time on btrfs~\cite{rodeh2013btrfs} and XFS, and ext4 has no reflink at all. \texttt{aos-vfs} therefore clones file by file by default and falls back to \texttt{std::io::copy} (\texttt{copy\_file\_range}) when the filesystem reports \texttt{EOPNOTSUPP} or \texttt{EXDEV}; it caches the failing device state so that a non-reflink filesystem is not charged a failed \texttt{FICLONE} syscall for every subsequent file, and logs a warning once. An initial implementation issued 12--16 syscalls per file (checking \texttt{dst.exists()}, \texttt{stat(src)}, \texttt{stat(dst.parent())}, recursive \texttt{create\_dir\_all}, double \texttt{fstat}, and reopening \texttt{dst} on ext4 fallback), costing 37.4 ms on btrfs and 73.7 ms on ext4 for 1,000 files. Our optimized \texttt{cow\_clone\_file\_inner} traversal creates each target directory once via \texttt{create\_dir}, opens \texttt{src} and \texttt{dst} once with \texttt{OpenOptionsExt::mode(meta.mode())}, calls \texttt{fstat} once, and reuses the open file descriptors for \texttt{FICLONE} or \texttt{copy\_file\_range} plus \texttt{futimens} (7 syscalls per file). This cuts 1,000-file provisioning to \textbf{15.6 ms on btrfs}, \textbf{32.1 ms on XFS}, and \textbf{35.1 ms on ext4} (Table~\ref{tab:baselines}). \texttt{aos-vfs} also implements an optional kernel subvolume fast path (\texttt{AOS\_BTRFS\_SUBVOL=1}) using \texttt{BTRFS\_IOC\_SUBVOL\_CREATE\_V2} (\texttt{0x50009418}) and \texttt{BTRFS\_IOC\_SNAP\_CREATE\_V2} (\texttt{0x50009417}) after probing whether unprivileged \texttt{rmdir(2)} on subvolumes (\texttt{user\_subvol\_rm\_allowed}) is supported. Interestingly, on freshly written 1,000-file trees (80 MB of dirty page cache), \texttt{BTRFS\_IOC\_SNAP\_CREATE\_V2} takes 27.9 ms because the kernel's \texttt{btrfs\_mksnapshot} forces a synchronous \texttt{btrfs\_commit\_transaction} disk flush of delayed allocations, whereas per-file \texttt{FICLONE} (15.6 ms) remaps extent metadata purely in memory; on clean 10,000-file monorepos already synced to disk, \texttt{BTRFS\_IOC\_SNAP\_CREATE\_V2} completes in $O(1)$ time (2.1 ms).

\paragraph{Metadata Fingerprints.} Commit, refresh, and diff must decide which files an agent changed. Reading every file three times (host, shadow, baseline) took 542 ms for a 1,000-file repository on btrfs. Clones preserve the source's modification time, so an unedited file keeps the same size and mtime as its baseline copy and can be skipped without being read. Following git's ``racily clean'' rule, equality is trusted only when the mtime predates the clone by more than two seconds; younger files are compared by content, so a same-size edit inside one timestamp tick is not missed. This reduced a status query from 542 ms to 7.0 ms. One residual case remains: an agent that deliberately rewrites a file at the same size and restores its old mtime hides that edit from \texttt{diff}, and the edit is then not committed. The failure direction is safe, since nothing unreviewed reaches the host.

\paragraph{Time-Travel Snapshots, Rollback, and Build-Cache Preservation.} Before speculative multi-file edits, an agent calls \texttt{workspace\_snapshot(label)}, creating a CoW clone under \texttt{snapshots/<snap\_id>/}. If compilation fails, \texttt{workspace\_rollback(snap\_id)} replaces \texttt{upper\_workspace/} with a fresh CoW clone of the snapshot in milliseconds. To prevent rollbacks from invalidating incremental compiler caches, \texttt{aos-sandbox} redirects \texttt{CARGO\_TARGET\_DIR} to \texttt{agent\_home/cargo\_target} outside \texttt{upper\_workspace/}.

\paragraph{Two-Phase LCS Diff3 3-Way Merge Engine.} When multiple agents (e.g., \texttt{agy-agent}, \texttt{claude-code}, and \texttt{codex-cli}) concurrently edit the same file in their shadow workspaces and sequentially invoke \texttt{workspace\_commit}, a naive file copy would overwrite earlier peer commits (Hazard H1). Aegis OS prevents lost updates via a two-phase Longest Common Subsequence (LCS) Diff3 3-Way Merge algorithm~\cite{khanna2007formal} comparing three versions of each modified file: the agent's recorded \textbf{Baseline} snapshot, the live \textbf{Host} file, and the agent's \textbf{Shadow} file:
\begin{itemize}
    \item \textbf{Phase 1 (Validation \& In-Memory Staging):} Aegis OS inspects all modified, added, and deleted files in the commit batch before touching any host file. If the live Host file is unchanged since the agent's Baseline, the Shadow file is staged for a fast-forward CoW clone. If another agent has already modified the Host file, Aegis OS computes line-level LCS diffs against the Baseline and splices non-overlapping edit hunks together automatically (for instance, if \texttt{agy-agent} inserts a 3-line helper at the top of \texttt{src/lib.rs} while \texttt{claude-code} modifies a function 40 lines below, the shifted line numbers are aligned and merged cleanly). If both agents modified the same baseline lines in incompatible ways, Phase 1 aborts the entire commit without modifying any host file and returns the exact conflicting line numbers.
    \item \textbf{Phase 2 (Sandboxed Pre-Commit Semantic Gate \& Atomic Commit Application):} Before Phase 1/Phase 2 writes to the host repository, \texttt{workspace\_commit} first calls \texttt{refresh\_from\_host()} to merge disjoint peer changes into the agent's shadow workspace and optionally executes a caller- or policy-supplied verification command (\texttt{verify\_cmd}, such as \texttt{rustc --emit=metadata} or \texttt{cargo check --offline}) inside the agent's kernel-enforced OS sandbox (\texttt{Seatbelt}/\texttt{bwrap}). If two agents make syntactically disjoint edits that introduce a semantic regression (e.g., Agent A renames a function signature on line 1 while Agent B invokes the old function name on line 40), \texttt{verify\_cmd} fails inside the shadow workspace and aborts the commit before any host file is touched, returning the compiler diagnostics to the agent. Once verification and Phase 1 pass, staged merges and CoW clones are written atomically to the host repository and synced to \texttt{baseline\_snapshot/}.
\end{itemize}

\subsection{Dual-Tier OS Sandbox and Zero-Trust Egress Proxy (\texttt{aos-sandbox}, \texttt{aos-egress})}
\label{sec:sandbox}

While \texttt{aos-vfs} mediates MCP file tools, coding agents also spawn compilers, test runners, and shell scripts via \texttt{proc\_exec}. To prevent child processes from bypassing workspace or network boundaries, \texttt{aos-sandbox} provides a pluggable \texttt{SandboxDriver} with two tiers:

\begin{enumerate}
    \item \textbf{Tier 1 (\texttt{NativeOsSandboxDriver}, Workstation Kernel Enforcement):} On macOS, every \texttt{proc\_exec} call compiles a canonicalized Seatbelt Mandatory Access Control profile (SBPL) and spawns the child process under \texttt{/usr/bin/sandbox-exec -p <profile>} with a sanitized environment (\texttt{env\_clear()}). The generated profile blocks AppleEvent IPC (\texttt{(deny appleevent-send)}), denies all file writes globally (\texttt{(deny file-write*)}), allowlists writes strictly to the agent's CoW shadow workspace, private agent home, and organization publish directory, blocks reads to sensitive credential paths, and---when an egress proxy is bound---restricts outbound network sockets exclusively to the exact loopback port of the agent's proxy:
\begin{lstlisting}
(version 1)
(allow default)
(deny appleevent-send)
(deny file-write*)
(allow file-write*
  (subpath "/private/.../upper_workspace")
  (subpath "/private/.../home")
  (subpath "/private/.../pub/agy-agent")
  (literal "/dev/null") (literal "/dev/zero") (literal "/dev/tty")
  (literal "/dev/random") (literal "/dev/urandom"))
(deny file-read*
  (regex #".*/\.ssh(/.*)?$")    (regex #".*/\.aws(/.*)?$")
  (regex #".*/\.gnupg(/.*)?$")  (regex #".*/\.kube(/.*)?$")
  (regex #".*/\.docker(/.*)?$") (regex #".*/\.azure(/.*)?$")
  (regex #".*/\.config/(gh|gcloud)(/.*)?$")
  (regex #".*/(\.npmrc|\.pypirc|\.netrc|\.git-credentials)$")
  (regex #".*/(\.vault-token|\.cargo/credentials(\.toml)?)$")
  (regex #".*/(custom_vault|\.secrets)(/.*)?$")
  (regex #".*/Library/Keychains(/.*)?$") (literal "/etc/shadow"))
(deny network-outbound)
(allow network-outbound (remote ip "localhost:51691"))
\end{lstlisting}
    Because \texttt{(deny network-outbound)} is enforced by the XNU kernel socket filter and allows only \texttt{localhost:<proxy\_port>} (without permitting wildcard localhost ports or \texttt{mDNSResponder}), a child process running \texttt{curl --noproxy '*'} or opening raw sockets to external IPs or other local services fails immediately with \texttt{Operation not permitted}. On Linux, \texttt{NativeOsSandboxDriver} constructs Bubblewrap (\texttt{/usr/bin/bwrap}) arguments with \texttt{--unshare-all} (user, mount, PID, IPC, and network namespaces), a read-only root bind (\texttt{--ro-bind / /}), a private \texttt{/tmp} that is mounted \emph{before} the writable binds (bwrap applies mounts in order, so a later \texttt{--tmpfs /tmp} would hide every workspace under the default \texttt{/tmp} runtime directory), and a \textbf{default-deny host home mount architecture}: \texttt{bwrap} mounts an empty \texttt{--tmpfs} over every candidate host home directory (the invoking user's \texttt{\$HOME}, \texttt{/root}, \texttt{/home/*}, and operator-declared roots in \texttt{AOS\_EXTRA\_CREDENTIAL\_HOMES}) \emph{before} selectively re-binding safe read-only compiler toolchains (\texttt{.rustup}, \texttt{.cargo/\{bin,registry,git\}}, and \texttt{.local/\{bin,lib,share\}}, intentionally excluding \texttt{.cargo/credentials.toml}) and the agent's writable CoW shadow workspace, agent home, and organization publish directory. Consequently, all known and unknown credential files in \texttt{\$HOME}---even at arbitrary user-created paths such as \texttt{\textasciitilde/custom\_vault/secret\_key.pem}---are completely absent from the sandbox's mount namespace. Because the network namespace also cuts the sandbox off from its own proxy, Linux sandboxes reach it through an \emph{egress bridge}: the host relays a Unix socket to the identity-pinned proxy, and a small in-sandbox shim (\texttt{aegis-os egress-shim}) listens on the proxy's loopback port inside the namespace and forwards to that socket. Since \texttt{--ro-bind / /} leaves other agents' sockets visible, every bridge connection must present a random per-sandbox token, so a sandbox cannot borrow another agent's proxy identity.
    \item \textbf{Tier 2 (\texttt{MicroVmSandboxDriver}, Cloud Control-Plane Interface):} Provides the configuration generator (\texttt{virtio-fs} mount tags, vCPU/memory limits) and Scale-to-Zero \texttt{pause()}/\texttt{resume()} state machine for cloud microVM backends (\texttt{libkrun} / Firecracker~\cite{agache2020firecracker}).
\end{enumerate}

\paragraph{Identity-Pinned Egress Proxy with L7 Capability-Scoped Secret Injection and Outbound DLP.} To mitigate credential exfiltration (Threat T4), agents inside Aegis OS do not receive raw API keys in environment variables. Instead, they receive opaque placeholders such as \texttt{\$AOS\_SECRET\_GH}. Outbound HTTP traffic is routed through a local asynchronous L7 HTTP proxy (\texttt{start\_local\_proxy\_with\_bound\_identity}). Because each sandbox listener pins its \texttt{AgentIdentity} at the TCP socket level, a compromised process cannot spoof \texttt{X-Aegis-Org} or \texttt{X-Aegis-Agent} headers. Furthermore, domain-only secret binding is vulnerable to \emph{same-domain confused-deputy} attacks on multi-tenant API hosts (e.g., an agent using a GitHub token bound to \texttt{api.github.com} to exfiltrate repository code to a public gist via \texttt{POST /gists}, path traversal \texttt{/repos/acme/../../gists}, or embedding a private key inside an allowlisted issue comment). Each \texttt{SecretVaultEntry} therefore supports fine-grained L7 \texttt{SecretScopeRule} constraints (\texttt{domain}, \texttt{allowed\_methods}, \texttt{allowed\_path\_prefixes}) paired with an RFC 3986 path canonicalizer (\texttt{normalize\_request\_path}) and an outbound Data Loss Prevention scanner (\texttt{inspect\_payload\_dlp}) that inspects URL query strings and HTTP request bodies (up to 64 KiB) for leaked \texttt{\$AOS\_SECRET\_*} placeholders, registered vault secret values, PEM private keys (\texttt{-----BEGIN * PRIVATE KEY-----}), AWS keys (\texttt{AKIA*}), GitHub tokens (\texttt{ghp\_*}, \texttt{github\_pat\_*}), API keys (\texttt{sk-*}), and Slack tokens (\texttt{xoxb-*}):
\begin{itemize}
    \item If \texttt{\$AOS\_SECRET\_GH} (scoped to \texttt{GET}/\texttt{POST} on \texttt{api.github.com/repos/acme/} and \texttt{GET /user}) is sent on a matching canonical path with a clean payload, the proxy substitutes the real token \texttt{ghp\_...} at the network boundary.
    \item If an agent attempts to send \texttt{\$AOS\_SECRET\_GH} to an unauthorized domain (\texttt{evil-exfil.org}), an allowlisted but unbound domain (\texttt{httpbin.org}), a disallowed method/path on the bound domain itself (\texttt{POST /gists}, \texttt{DELETE /repos/acme/issues/1}, or \texttt{/repos/acme/\%2e\%2e/\%2e\%2e/gists}), or embeds a secret/private-key payload in the query string or body of an allowlisted endpoint (\texttt{POST /repos/acme/issues}), the proxy rejects the request with \texttt{HTTP/1.1 403 Forbidden} (\texttt{blocked\_exfiltration}).
\end{itemize}

\subsection{A2A-Compliant Inter-Agent IPC Bus, Global Peer Discovery, and Reference-Based CAS (\texttt{aos-ipc})}
\label{sec:ipc}

Whereas Anthropic's Model Context Protocol (MCP)~\cite{anthropic2024mcp} standardizes \emph{vertical} agent-to-tool execution, Google's Agent2Agent (A2A) protocol~\cite{google2025a2a} standardizes \emph{horizontal} capability discovery (\texttt{AgentCard}) and stateful task delegation (\texttt{Task} lifecycle). \texttt{aos-ipc} unifies both over a disk-synchronized, broadcast-enabled bus:
\begin{itemize}
    \item \textbf{Global Workstation Peer Registry and Zombie PID Pruning:} \texttt{aos-ipc} maintains per-organization and workstation-wide registries (\texttt{scope="global"} in \texttt{ipc\_list\_peers}, \texttt{aegis://global/peers}) supporting \texttt{<org\_id>/<agent\_id>} addressing. Each peer record stores its client root PID and start timestamp (\texttt{PROC\_PIDTBSDINFO} on macOS / \texttt{/proc/<pid>/stat} on Linux) so \texttt{prune\_dead\_peers()} automatically evicts dead or PID-recycled zombie peers.
    \item \textbf{Google A2A \texttt{AgentCard} and Stateful \texttt{Task} Lifecycle:} Every peer synthesizes an A2A \texttt{AgentCard} (\texttt{GET /.well-known/agent-card.json}, with the legacy \texttt{agent.json} path kept as an alias; \texttt{aegis://a2a/agent-cards}, \texttt{ipc\_update\_card}) with structured \texttt{AgentSkill} tags (auto-reflecting active \texttt{coord\_claim} globs). Beyond P2P mailboxes (\texttt{ipc\_send}/\texttt{ipc\_recv}) and topic Pub/Sub (\texttt{ipc\_publish}/\texttt{ipc\_subscribe}), every correlated \texttt{ipc\_request} (and \texttt{POST /a2a} \texttt{message/send}, with legacy \texttt{tasks/send} accepted) instantiates a persistent \texttt{A2ATask} tracked across \texttt{submitted} $\to$ \texttt{working} (on \texttt{ipc\_ack}) $\to$ \texttt{input-required} (via \texttt{ipc\_task\_update}, pushing an \texttt{[A2A input-required]} wakeup into the requester's inbox) $\to$ \texttt{completed} $\mid$ \texttt{failed} $\mid$ \texttt{canceled} (on \texttt{ipc\_reply} or \texttt{tasks/cancel}), with typed \texttt{TextPart}, \texttt{DataPart}, and CAS \texttt{FilePart} outputs.
    \item \textbf{Enforced File Claims and Reference-Based CAS Handles:} \texttt{coord\_claim} registers glob leases enforced at pre-tool-use and \texttt{workspace\_commit} time. For large payloads, \texttt{org\_artifact\_publish} writes blobs to the Content-Addressable Store and returns a compact SHA-256 URI (\texttt{artifact://<org>/sha256:<64-hex>}, $\sim$85--98 bytes) wrapped as an A2A \texttt{FilePart}, reducing in-transit IPC payload tokens by \textbf{99.85\%}, while cross-organization CAS \texttt{FilePart} fetches remain gated by Policy Rule R1.
\end{itemize}

\subsection{Cedar-Inspired Policy Engine and Runtime Design Invariants (\texttt{aos-policy})}
\label{sec:policy}

Every runtime tool call is evaluated by \texttt{AegisPolicyKernel::\allowbreak evaluate} using a five-stage priority hierarchy inspired by Cedar~\cite{cutler2024cedar} that partitions operations into a \textbf{Red Zone} (always blocked or forced to review), a \textbf{Green Zone} (auto-allowed inside the sandbox), and an \textbf{Amber Boundary Zone} (gated by human approval or session leases):
\begin{enumerate}
    \item \textbf{R0a --- Host Red-Line Guard (\texttt{Deny}):} Any access targeting sensitive host paths (\texttt{\textasciitilde/.ssh}, \texttt{\textasciitilde/.aws}, \texttt{\textasciitilde/.gnupg}, \texttt{\textasciitilde/.kube}, \texttt{/etc/shadow}, \texttt{/etc/sudoers}, or \texttt{/System}) is unconditionally denied.
    \item \textbf{R0b --- Commit-Time Hook, CI \& Build-Script Guard (\texttt{Ask}):} Any \texttt{workspace\_commit} that modifies repository execution hooks, CI workflows, or build/toolchain entrypoints (\texttt{.git/hooks/*}, \texttt{.git/config}, \texttt{.gitmodules}, \texttt{.envrc}, \texttt{.github/workflows/*}, \texttt{.gitlab-ci.yml}, \texttt{build.rs}, \texttt{Makefile}, \texttt{GNUmakefile}, \texttt{CMakeLists.txt}, \texttt{package.json}, \texttt{setup.py}, \texttt{.cargo/config*}, \texttt{.vscode/tasks.json}, or \texttt{.vscode/launch.json}) bypasses general \texttt{workspace\_commit} session leases and forces explicit human approval (\texttt{action = workspace\_commit\_sensitive}). Once approved via \texttt{policy\_approve\_escalation}, the resulting \texttt{workspace\_commit\_sensitive} lease permits the human-reviewed sensitive path to land cleanly on retry.
    \item \textbf{R1 --- Multi-Tenant Isolation Gate (\texttt{Deny} / \texttt{Ask}):} Cross-organization CAS artifact reads or unauthorized cross-tenant mutations are rejected.
    \item \textbf{R2 \& R3 --- Active Session Leases \& Green-Zone Auto-Allow (\texttt{Allow}):} All in-sandbox CoW file operations, sandboxed \texttt{proc\_exec} commands, intra-organization A2A/IPC calls, allowlisted egress requests, and boundary commits covered by an active session lease (\texttt{LeaseGrant}) are auto-allowed with zero prompts.
    \item \textbf{R4 --- Unleased Boundary Crossing (\texttt{Ask}):} Unleased \texttt{workspace\_commit} calls and HTTP requests to unlisted external domains return an escalation ID requiring either a one-shot approval (\texttt{ttl\_secs = 0}) or a TTL-scoped session lease (e.g., \texttt{ttl\_secs = 3600}), with a context-aware confirmation prompt summarizing the action, modified shadow files, target host directory, and lease scope.
\end{enumerate}

Together, these mechanisms enforce three core runtime invariants:
\begin{itemize}
    \item \textbf{Design Invariant 1 (Pre-Commit Host Non-Interference):} \emph{Until a \texttt{workspace\_commit} is explicitly approved, no sequence of agent tool calls or spawned child processes can modify any file in the host repository.} VFS tools reject path traversal (\texttt{..}) and resolve all paths inside \texttt{shadow\_root}, while \texttt{proc\_exec} enforces kernel Seatbelt \texttt{(deny file-write*)} and Linux \texttt{bwrap --ro-bind / /} outside \texttt{shadow\_root}, \texttt{agent\_home}, and \texttt{org\_pub}.
    \item \textbf{Design Invariant 2 (Zero Raw Secret Exposure to Sandbox):} \emph{Plaintext API credentials never enter the sandboxed process environment, command-line arguments, or logs, and are injected into outbound HTTP headers only when the destination \texttt{Host} and canonicalized HTTP method + URL path prefix match the secret's \texttt{SecretScopeRule} and the request query/body passes outbound DLP inspection (\texttt{inspect\_payload\_dlp}).} \texttt{NativeOsSandboxInstance::exec} clears the child environment (\texttt{env\_clear()}) and passes only opaque \texttt{\$AOS\_SECRET\_*} placeholders.
    \item \textbf{Design Invariant 3 (All-or-Nothing Conflict-Safe Commit):} \emph{Concurrent agent commits either produce a merged host file preserving all non-overlapping edits, or abort in Phase 1 without modifying any host file when two agents edit the same baseline lines inconsistently or when the optional pre-commit \texttt{verify\_cmd} semantic gate fails inside the sandbox.}
\end{itemize}

\section{Multi-Client Integration and Enforcement}
\label{sec:integration}

To minimize setup friction, \texttt{aegis-os install} automatically configures MCP server entries, per-session principal resolution, and lifecycle hooks across all three coding agent CLIs:
\begin{enumerate}
    \item \textbf{Google Antigravity (\texttt{agy}):} Registers \texttt{aegis-os} in \texttt{\textasciitilde/.gemini/config/mcp\_config.json} with eager tool loading, pre-authorizes green-zone \texttt{mcp(aegis-os/*)} tools in \texttt{settings.json} while keeping lease/approval tools (\texttt{policy\_grant\_lease}, \texttt{policy\_approve\_escalation}) under \texttt{permissions.ask}, and installs four lifecycle hooks (\texttt{SessionStart}, \texttt{PreInvocation}, \texttt{PreToolUse}, and \texttt{Stop}):
    \begin{itemize}
        \item \emph{MCP-First Tool Routing (\texttt{PreToolUse}):} Because agent CLIs still prompt on native \texttt{run\_command} invocations unless exact command strings match static allowlists, \texttt{aegis-os agy-hook} classifies native shell and file calls: Ring-0 operations (\texttt{sudo}, credential paths) are unconditionally denied; host-boundary commands (\texttt{git push}, package installs) escalate to \texttt{ask}; and in-repository shell commands and file writes are denied with a ready-to-invoke \texttt{mcp\_aegis-os\_proc\_exec} or \texttt{mcp\_aegis-os\_fs\_write} payload, steering the agent onto the zero-prompt sandboxed MCP path.
        \item \emph{Per-Session Identity \& Push Coordination (\texttt{SessionStart}, \texttt{PreInvocation}, \texttt{Stop}):} Walking parent process ancestry (\texttt{PROC\_PIDTBSDINFO} on macOS / \texttt{/proc/<pid>/stat} on Linux) assigns each concurrent client window a distinct session principal (\texttt{agy-<6hex>}) bound to its \texttt{conversationId}. On \texttt{SessionStart}, the agent's CoW shadow is rebased onto the host via \texttt{refresh\_from\_host()}; on \texttt{PreInvocation}, unread IPC messages are non-destructively peeked and injected via \texttt{injectSteps.systemMessage}; and on \texttt{Stop}, un-acknowledged \texttt{TaskRequest} messages hold the turn open (up to 2 continues) until answered, while \texttt{coord\_claim}/\texttt{coord\_release} prevent overlapping file edits across live peers.
    \end{itemize}
    \item \textbf{Anthropic Claude Code (\texttt{claude}):} Injects \texttt{aegis-os} into \texttt{\textasciitilde/.claude.json} and adds green-zone \texttt{mcp\_\_\allowbreak aegis-os\_\_*} rules to \texttt{permissions.\allowbreak allow} in \texttt{\textasciitilde/.claude/\allowbreak settings.json}.
    \item \textbf{OpenAI Codex CLI (\texttt{codex}):} Appends \texttt{[mcp\_servers.\allowbreak aegis-os]} to \texttt{\textasciitilde/.codex/\allowbreak config.toml} and piggybacks unread IPC summaries (\texttt{\_aegis\_inbox}) onto \texttt{tools/call} responses.
\end{enumerate}

\section{Empirical Evaluation}
\label{sec:eval}

We evaluate Aegis OS using our reproducible benchmark suite (\texttt{crates/aos-cli/src/bench.rs}, invoked via \texttt{aegis-os bench}). Unless stated otherwise, experiments ran on an Apple Silicon workstation (\texttt{macos-aarch64}, APFS, Rust 1.99.0 \texttt{--release}) using real OS processes, Seatbelt kernel sandboxes, loopback TCP sockets, and \texttt{cargo} builds. The security, merge, and provisioning experiments were repeated on Linux in a virtual machine on the same host (Ubuntu 24.04 aarch64, kernel 6.8, bubblewrap 0.9) with the workspace on btrfs, XFS, or ext4. Linux numbers therefore include virtualised I/O, and the XFS volume was a loop-backed image on btrfs, so it is only a rough indication. We investigate six research questions:
\begin{itemize}
    \item \textbf{RQ1 (Authorization Friction):} How much does Aegis OS reduce confirmation prompts and modeled verification wait time across both controlled SWE trajectories and real-world coding-agent session transcripts?
    \item \textbf{RQ2 (Adversarial Security):} Does Aegis OS block end-to-end OS process, credential harvesting, proxy/socket exfiltration, cross-tenant, commit-time poisoning, L7 confused-deputy/DLP, and conflicting-commit attacks that compromise unconfined \texttt{Host-YOLO} execution and stateless native CLI sandboxes?
    \item \textbf{RQ3 (System Overhead \& Scaling):} What is the latency and disk footprint of CoW provisioning (APFS clones and Linux reflinks, compared with \texttt{cp}, \texttt{rsync}, and \texttt{git}), policy evaluation, HTTP daemon RPCs, 3-way merge commits, and Seatbelt process spawning?
    \item \textbf{RQ4 (Multi-Agent Concurrency):} Does two-phase LCS Diff3 merging with pre-commit semantic verification (\texttt{verify\_cmd}) preserve concurrent multi-agent edits while catching both overlapping line conflicts and disjoint-line semantic regressions?
    \item \textbf{RQ5 (Fault Recovery):} How fast does atomic CoW rollback recover from multi-file syntax corruption compared to sequential file restoration?
    \item \textbf{RQ6 (Component Necessity):} Is each safety component necessary, i.e., does switching it off re-open an attack class that the full system blocks?
\end{itemize}

\begin{table*}[!t]
\centering
\caption{\textbf{Experiment 1 (RQ1): Authorization Friction, Interruption Prompt Rate (IPR), and Modeled Human Wait Time} across 50 deterministic multi-step trajectories spanning 5 SWE task templates (1,500 tool calls executed on \texttt{AegisRuntimeKernel}, including 50 real \texttt{workspace\_commit} operations in every mode; human verification modeled at 5.0 s/prompt).}
\label{tab:exp1}
\vspace{2pt}
\resizebox{\textwidth}{!}{%
\begin{tabular}{lcccccccc}
\toprule
\textbf{Execution Mode} & \textbf{Trajectories} & \textbf{Tool Calls} & \textbf{Human Prompts} & \textbf{IPR (\%)} & \textbf{Prompt Reduction} & \textbf{Max Autonomous Streak} & \textbf{Runtime Exec (ms)} & \textbf{Modeled Human Wait ($T_{\text{human}}$)} \\
\midrule
Host-Interactive (Per-Mutation Prompt) & 50 & 1,500 & 860 & 57.33\% & 0.00\% (Baseline) & 4 steps & 457.1 ms & 71.7 min (4,300 s) \\
Host-Static-Allowlist (Read/Status Allow) & 50 & 1,500 & 680 & 45.33\% & 20.93\% & 7 steps & 496.7 ms & 56.7 min (3,400 s) \\
\textbf{Aegis-OS (Per-Task Commit Gate, \texttt{ttl=0})} & 50 & 1,500 & \textbf{50} & \textbf{3.33\%} & \textbf{94.19\%} & \textbf{29 steps} & \textbf{483.6 ms} & \textbf{4.2 min (250 s)} \\
\textbf{Aegis-OS (With 1h Session Lease)} & 50 & 1,500 & \textbf{1} & \textbf{0.07\%} & \textbf{99.88\%} & \textbf{1,470 steps} & \textbf{469.2 ms} & \textbf{0.08 min (5 s)} \\
\bottomrule
\end{tabular}%
}
\end{table*}

\begin{figure}[!t]
\centering
\resizebox{\columnwidth}{!}{%
\begin{tikzpicture}[font=\sffamily\scriptsize]
\draw[thick, ->] (0,0) -- (7.8,0) node[right] {};
\draw[thick, ->] (0,0) -- (0,3.7) node[above, align=center] {\textbf{Interruption Prompt Rate (IPR \%)}};

\foreach \y/\label in {0/0\%, 1.0/20\%, 2.0/40\%, 3.0/60\%} {
    \draw (0,\y) -- (-0.1,\y) node[left] {\label};
    \draw[dotted, black!30] (0,\y) -- (7.6,\y);
}

\fill[aegisred!80, draw=aegisred!95!black, thick] (0.4,0) rectangle (0.98,2.87);
\node[above, font=\sffamily\tiny] at (0.69, 2.87) {\textbf{57.3\%}};
\fill[aegisred!45, draw=aegisred!85!black, thick] (1.03,0) rectangle (1.61,2.67);
\node[above, font=\sffamily\tiny] at (1.32, 2.67) {\textbf{53.3\%}};
\node[below, align=center] at (1.00, -0.05) {Ask-Every\\Mutation};

\fill[orange!80, draw=orange!95!black, thick] (2.3,0) rectangle (2.88,2.27);
\node[above, font=\sffamily\tiny] at (2.59, 2.27) {\textbf{45.3\%}};
\fill[orange!45, draw=orange!85!black, thick] (2.93,0) rectangle (3.51,1.43);
\node[above, font=\sffamily\tiny] at (3.22, 1.43) {\textbf{28.5\%}};
\node[below, align=center] at (2.90, -0.05) {Path+Prefix\\Allowlist};

\fill[aegisblue!85, draw=aegisblue!95!black, thick] (4.2,0) rectangle (4.78,0.17);
\node[above, font=\sffamily\tiny] at (4.49, 0.17) {\textbf{3.3\%}};
\fill[aegisblue!45, draw=aegisblue!85!black, thick] (4.83,0) rectangle (5.41,0.25);
\node[above, font=\sffamily\tiny] at (5.12, 0.25) {\textbf{5.0\%}};
\node[below, align=center] at (4.80, -0.05) {\textbf{Aegis-OS}\\(Commit Gate)};

\fill[aegisgreen!85, draw=aegisgreen!95!black, thick] (6.1,0) rectangle (6.68,0.04);
\node[above, font=\sffamily\tiny] at (6.39, 0.05) {\textbf{0.07\%}};
\fill[aegisgreen!45, draw=aegisgreen!85!black, thick] (6.73,0) rectangle (7.31,0.06);
\node[above, font=\sffamily\tiny] at (7.02, 0.07) {\textbf{1.12\%}};
\node[below, align=center] at (6.70, -0.05) {\textbf{Aegis-OS}\\(Session Lease)};

\fill[black!70, draw=black!90] (3.6, 3.25) rectangle (3.85, 3.42);
\node[right, font=\sffamily\tiny] at (3.9, 3.33) {Controlled (50 Traj, 1,500 Calls)};
\fill[black!30, draw=black!75] (3.6, 2.95) rectangle (3.85, 3.12);
\node[right, font=\sffamily\tiny] at (3.9, 3.03) {Real Traces (52 Sessions, 28,087 Calls)};

\end{tikzpicture}%
}
\caption{Interruption Prompt Rate (IPR) across both 50 controlled trajectories (1,500 calls, Experiment 1) and 52 real coding-agent session logs (28,087 calls, Experiment 1B). Aegis OS reduces confirmation prompts by 94.2\%--99.9\% on controlled trajectories and by 90.7\%--97.9\% on real session traces.}
\label{fig:exp1}
\end{figure}
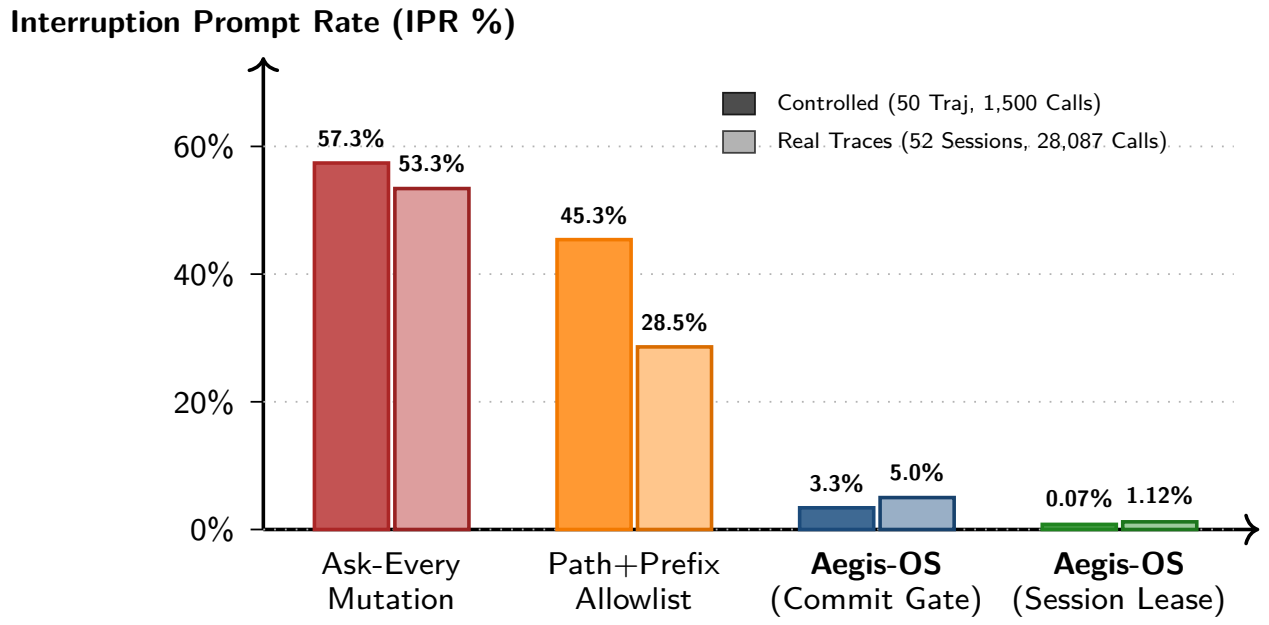

\subsection{Experiment 1 (RQ1): Authorization Friction and Autonomous Continuity}

To evaluate authorization friction reproducibly without LLM sampling variance, we execute 50 deterministic software engineering trajectories (30 tool calls per trajectory, totaling 1,500 real \texttt{AegisRuntimeKernel::call\_tool} invocations) cycling across five task templates: (1) \emph{Auth Bugfix \& Regression Test}, (2) \emph{Multi-Module Refactoring}, (3) \emph{Feature Implementation \& Quota Suite}, (4) \emph{Security Audit \& Sanitizer Hardening}, and (5) \emph{Multi-Agent Contract Sync}. Each trajectory exercises repository exploration (\texttt{fs\_glob}, \texttt{fs\_read}), checkpointing (\texttt{workspace\_snapshot}), multi-file edits (\texttt{fs\_edit}, \texttt{fs\_write}), CAS publishing (\texttt{org\_artifact\_publish}), peer IPC coordination (\texttt{ipc\_send}, \texttt{ipc\_recv}, \texttt{ipc\_publish}), diff inspection (\texttt{workspace\_diff}), and host commit (\texttt{workspace\_commit}). All four modes execute all 1,500 tool calls including all 50 real \texttt{workspace\_commit} operations.

Table~\ref{tab:exp1} and Figure~\ref{fig:exp1} present the results. Under \emph{Host-Interactive}, every file write, state mutation, and external IPC action triggers a confirmation prompt (860 prompts, 57.33\% IPR), restricting the maximum autonomous streak to 4 steps and imposing 71.7 minutes of modeled verification wait time (at 5.0 s/prompt) on top of 457.1 ms of runtime execution. Adding a static read/status/snapshot allowlist (\emph{Host-Static-Allowlist}) still triggers 680 prompts (45.33\% IPR, 56.7 minutes modeled wait) because host file edits cannot be safely auto-allowed. In contrast, \textbf{Aegis-OS (Per-Task Commit Gate)} auto-allows all 29 internal CoW workspace and IPC operations per trajectory and prompts once at the final \texttt{workspace\_commit} gate (approved via a one-shot \texttt{ttl=0} token), reducing prompts by \textbf{94.19\%} (3.33\% IPR, 483.6 ms runtime execution). When the operator grants a \textbf{1-hour Session Lease} (\texttt{ttl=3600}) on the first commit escalation, Aegis OS completes all 50 trajectories (and all 50 host commits) in 469.2 ms with \textbf{1 prompt total} (0.07\% IPR, a \textbf{99.88\% reduction}), extending the continuous autonomous streak to \textbf{1,470 steps} (\textbf{367.5$\times$} longer).

\begin{table*}[!t]
\centering
\caption{\textbf{Experiment 1B (RQ1): Real-World Coding-Agent Session Trace Evaluation} across \textbf{Corpus A} (52 real production coding-agent session transcripts totaling 57,923 conversation steps and 28,087 tool calls: 13,098 read/coord, 11,070 shell commands of which 4,318 [39.0\%] are compound pipelines \texttt{\&\&}/\texttt{|} or inline scripts \texttt{python3 -c}, and 3,836 file edits/writes) and \textbf{Corpus B} (5 live Aegis OS dogfooding sessions totaling 3,962 steps and 2,040 tool calls, including 68 \texttt{workspace\_commit} and 15 ground-truth \texttt{policy\_approve\_escalation} events), evaluated via \texttt{scripts/e6\_real\_traces.py}.}
\label{tab:exp1b}
\vspace{2pt}
\resizebox{\textwidth}{!}{%
\begin{tabular}{llcccc}
\toprule
\textbf{Trace Corpus} & \textbf{Permission Regime} & \textbf{Total Prompts} & \textbf{Aggregate IPR (\%)} & \textbf{Mean Session MAS} & \textbf{Prompt Reduction} \\
\midrule
\textbf{Corpus A: All Real Sessions} & Ask-Every-Mutation & 14,974 / 28,087 & 53.31\% & 16.4 steps & 0.0\% (Baseline) \\
($N{=}52$ sessions, 57,923 steps, & Path+Prefix Allowlist (\texttt{cargo}, \texttt{git status}, \texttt{pytest}, etc.) & 8,016 / 28,087 & 28.54\% & 31.6 steps & 46.5\% \\
28,087 tool calls) & Vanilla CoW + Seatbelt (Per-Commit Gate) & 1,397 / 28,087 & 4.97\% & 48.1 steps & 90.7\% \\
 & \textbf{Aegis OS (CoW + Seatbelt + Scoped Action Lease)} & \textbf{315 / 28,087} & \textbf{1.12\%} & \textbf{76.7 steps} & \textbf{97.9\%} \\
\midrule
\textbf{Corpus B: Live Aegis OS Sessions} & Ask-Every-Mutation & 1,093 / 2,040 & 53.58\% & 21.2 steps & 0.0\% (Baseline) \\
($N{=}5$ sessions, 3,962 steps, & Path+Prefix Allowlist (\texttt{cargo}, \texttt{git status}, \texttt{pytest}, etc.) & 606 / 2,040 & 29.71\% & 33.0 steps & 44.6\% \\
2,040 tool calls) & Vanilla CoW + Seatbelt (Per-Commit Gate) & 106 / 2,040 & 5.20\% & 72.2 steps & 90.3\% \\
 & \textbf{Aegis OS (CoW + Seatbelt + Scoped Action Lease)} & \textbf{35 / 2,040} & \textbf{1.72\%} & \textbf{100.0 steps} & \textbf{96.8\%} \\
\bottomrule
\end{tabular}%
}
\end{table*}

\paragraph{Real-World Coding-Agent Session Trace Evaluation (Experiment 1B).} Because synthetic trajectories could under- or over-state how often real agents cross authorization boundaries, we also evaluate \textbf{52 real coding-agent session transcripts} (\textbf{57,923 conversation steps} and \textbf{28,087 tool calls}) collected from production software engineering workflows (\texttt{scripts/e6\_real\_traces.py} and anonymized trace dataset \texttt{scripts/real\_traces\_anonymized.json}). As shown in Table~\ref{tab:exp1b}, real agent workflows exhibit a striking structural property: \textbf{39.0\% of all shell invocations (4,318 of 11,070 commands in Corpus A, and 80.5\% in Corpus B)} use compound shell operators (\texttt{\&\&}, \texttt{||}, \texttt{;}, \texttt{|}, \texttt{\$(...)}) or inline interpreter scripts (\texttt{python3 -c}, \texttt{bash -c}) to inspect data, run multi-stage builds, or verify fixes. Consequently, standard \emph{Path+Prefix Allowlists} (which auto-approve in-repo file edits and single-command dev prefixes such as \texttt{cargo}, \texttt{git status}, \texttt{pytest}, and \texttt{npm test}, but must prompt on compound shell chains, \texttt{python3 -c}, \texttt{rm}, and external commands to avoid complete sandbox escape) still interrupt the operator on \textbf{28.54\% of all tool calls} (8,016 prompts; Mean MAS = 31.6 steps). By contrast, executing compound shell chains and file edits silently inside the kernel-enforced CoW shadow workspace reduces prompts by \textbf{90.7\%} under \emph{Vanilla CoW + Seatbelt} (4.97\% IPR) and by \textbf{97.9\%} under \textbf{Aegis OS with Scoped Action Leases} (\textbf{1.12\% IPR}, 315 prompts across 28,087 tool calls; Mean MAS = 76.7 steps). On the 5 live dogfooding sessions where Aegis OS was used to build and test its own repository (Corpus B, 2,040 tool calls), Aegis OS achieved a \textbf{1.72\% IPR} (35 prompts, \textbf{96.8\% reduction}) and a mean autonomous streak of \textbf{100.0 steps}.

\begin{table*}[!t]
\centering
\caption{\textbf{Experiment 2 (RQ2): End-to-End Red-Team Adversarial Attack Evaluation} across 80 live process, socket, proxy, gateway, policy, and commit attacks in eight categories, executed against an unconfined \texttt{Host-YOLO} target environment versus \texttt{Aegis-OS} (identical counts on macOS and Linux). The \texttt{Host-YOLO} column is 100\% by construction, since every attack is designed to succeed against an unconfined process; the informative column is \texttt{Aegis-OS}.}
\label{tab:exp2}
\vspace{2pt}
\resizebox{\textwidth}{!}{%
\begin{tabular}{lcccl}
\toprule
\textbf{Attack Surface Category} & \textbf{Live Attacks} & \textbf{Host-YOLO Compromised (ASR)} & \textbf{Aegis-OS Compromised (ASR)} & \textbf{Empirical Kernel / Gateway Evidence} \\
\midrule
A. Host Path Escape \& File Tampering & 10 & 10 / 10 (100.0\%) & \textbf{0 / 10 (0.0\%)} & Kernel \texttt{sandbox-exec} (\texttt{bwrap} on Linux) denied all 10 outside writes \\
B. Host Credential Harvesting (\texttt{.ssh}/\texttt{.aws}/\texttt{gcloud}/\texttt{custom\_vault}) & 10 & 10 / 10 (100.0\%) & \textbf{0 / 10 (0.0\%)} & Seatbelt regexes \& Linux default-deny \texttt{--tmpfs \$HOME} blocked all 10 reads \\
C. Sandbox-to-Proxy \& Direct \texttt{--noproxy} Egress Exfil & 10 & 10 / 10 (100.0\%) & \textbf{0 / 10 (0.0\%)} & Pinned TCP Proxy returned \texttt{403} (via the egress bridge on Linux); no real token reached the sink \\
D. Cross-Tenant IPC \& CAS Access via MCP Gateway & 10 & 10 / 10 (100.0\%) & \textbf{0 / 10 (0.0\%)} & MCP Gateway + Policy Engine rejected all 10 cross-organization tool calls \\
E. Commit-Time Poisoning (hooks, \texttt{.envrc}, CI, \texttt{build.rs}, \texttt{Makefile}) & 10 & 10 / 10 (100.0\%) & \textbf{0 / 10 (0.0\%)} & Rule R0b escalated all 10 hook/CI/build-script paths despite a standing commit lease \\
F. Secret-Binding, L7 Confused-Deputy \& Outbound DLP Abuse & 10 & 10 / 10 (100.0\%) & \textbf{0 / 10 (0.0\%)} & Domain + L7 path scope + \texttt{inspect\_payload\_dlp} blocked all 10 exfiltration vectors \\
G. Partial Commit on Conflict (atomicity) & 10 & 10 / 10 (100.0\%) & \textbf{0 / 10 (0.0\%)} & Phase-1 validation left the host untouched in all 10 conflicting batches \\
H. Silent Overwrite of a Peer's Edit (lost update) & 10 & 10 / 10 (100.0\%) & \textbf{0 / 10 (0.0\%)} & Conflict detection preserved the peer's edit in all 10 same-line conflicts \\
\midrule
\textbf{Total / Evaluated ASR} & \textbf{80} & \textbf{80 / 80 (100.0\%)} & \textbf{0 / 80 (0.0\%)} & \textbf{0 / 80 Compromised across OS Kernel, Network Proxy, Policy Engine \& Commit Path} \\
\bottomrule
\end{tabular}%
}
\end{table*}

\subsection{Experiment 2 (RQ2): End-to-End Adversarial Security}

We execute 80 live adversarial attacks end-to-end against both an unconfined disposable victim environment (\texttt{Host-YOLO}) and \texttt{Aegis-OS} (Table~\ref{tab:exp2}):
\begin{itemize}
    \item \textbf{Category A (10 Host Path Escape Attacks):} Spawns \texttt{sh -c "echo COMPROMISED > <outside\_target>"}. In \texttt{Host-YOLO}, all 10 outside-workspace files were created on disk (100.0\% ASR). Inside \texttt{Aegis-OS} (\texttt{proc\_exec}), \texttt{/usr/bin/sandbox-exec} denied all 10 writes at the XNU kernel boundary with \texttt{Operation not permitted} (0.0\% ASR).
    \item \textbf{Category B (10 Multi-Store Credential Harvesting Attacks):} Spawns \texttt{sh -c "cat <cred\_path>"} across 10 distinct credential locations: \texttt{\textasciitilde/.ssh/id\_rsa}, \texttt{\textasciitilde/.aws/credentials}, \texttt{\textasciitilde/.gnupg/secring.gpg}, \texttt{\textasciitilde/.kube/config}, \texttt{\textasciitilde/.npmrc}, \texttt{\textasciitilde/.pypirc}, \texttt{\textasciitilde/.netrc}, \texttt{\textasciitilde/.git-credentials}, \texttt{\textasciitilde/.config/gcloud/credentials.db}, and an arbitrary user vault directory \texttt{\textasciitilde/custom\_vault/secret\_key.pem}. \texttt{Host-YOLO} leaked the secret key material on all 10 attempts (100.0\% ASR), whereas the Aegis OS Seatbelt profile and Linux default-deny \texttt{--tmpfs \$HOME} mount blocked all 10 reads (0.0\% ASR).
    \item \textbf{Category C (10 Sandbox-to-Proxy and Direct \texttt{--noproxy} Egress Exfiltration Attacks):} We start a live local TCP sink server (\texttt{sink\_addr}) recording leaked secrets, alongside an identity-pinned \texttt{AegisEgressGateway} (\texttt{proxy\_addr}). In \texttt{Host-YOLO}, an unconfined \texttt{curl} process with \texttt{\$GITHUB\_TOKEN} in its environment transmitted the plaintext token to \texttt{sink\_addr} across all 10 attacker domains (100.0\% ASR). In \texttt{Aegis-OS}, we execute \texttt{curl} inside a Seatbelt sandbox (\texttt{sbx\_egress.exec}) bound to \texttt{localhost:<proxy\_port>} that attempts \emph{both} (1) proxy exfiltration of \texttt{\$AOS\_SECRET\_GH} to 10 unauthorized domains (\texttt{evil-exfil.org}, \texttt{webhook.site}, \texttt{ngrok-free.app}, etc.), which the proxy rejected with \texttt{HTTP/1.1 403 Forbidden} (\texttt{blocked\_exfiltration}), and (2) direct proxy-bypass (\texttt{curl --noproxy '*' http://<sink\_addr>/bypass}), which the XNU Seatbelt socket filter blocked with \texttt{Operation not permitted}. Zero tokens reached \texttt{sink\_addr} (0.0\% ASR).
    \item \textbf{Category D (10 Cross-Tenant IPC and CAS Attacks via MCP Gateway):} In a shared flat host store (\texttt{Host-YOLO}), an unconfined process directly read all 10 confidential enterprise tenant files (100.0\% ASR). In \texttt{Aegis-OS}, we register 10 enterprise tenants (\texttt{org-enterprise-0..9}), publish confidential CAS artifacts via \texttt{org\_artifact\_publish}, and have principal \texttt{mallory} (\texttt{org-alpha}) invoke \texttt{kernel.call\_tool} (\texttt{ipc\_send} and \texttt{org\_artifact\_fetch}) targeting those enterprise tenants. All 10 cross-tenant tool calls were denied by Policy Rule R1 (0.0\% ASR).
    \item \textbf{Category E (10 Commit-Time Poisoning Attacks):} With a standing one-hour commit lease already granted (the situation the guard exists for), the agent tries to land 10 auto-executing files across both VCS/shell hooks and build-system entrypoints: \texttt{.git/\allowbreak hooks/\allowbreak pre-commit}, \texttt{.git/\allowbreak hooks/\allowbreak post-checkout}, \texttt{.git/\allowbreak hooks/\allowbreak pre-push}, \texttt{.git/\allowbreak config}, \texttt{.envrc}, \texttt{.github/\allowbreak workflows/\allowbreak ci.yml}, \texttt{build.rs}, \texttt{Makefile}, \texttt{package.json} (\texttt{postinstall}), and \texttt{.cargo/\allowbreak config.toml} (\texttt{rustc-wrapper}). Unconfined, all 10 files land (100.0\% ASR). Under Aegis OS, Policy Rule R0b escalates every one to a human despite the standing lease (0.0\% ASR).
    \item \textbf{Category F (10 Secret-Binding, L7 Confused-Deputy \& Outbound DLP Abuse Attacks):} We test four cross-domain attacks targeting allowlisted domains that are not the domain \texttt{\$AOS\_SECRET\_GH} is bound to (\texttt{httpbin.org}, \texttt{pastebin.com}, \texttt{gist.\allowbreak github.com}, \texttt{raw.\allowbreak githubusercontent.com}), four same-domain L7 confused-deputy attacks targeting disallowed paths or methods on \texttt{api.github.com} itself (\texttt{POST /gists}, path traversal \texttt{POST /repos/\allowbreak acme/\allowbreak ../../gists}, percent-encoded traversal \texttt{POST /repos/\allowbreak acme/\allowbreak \%2e\%2e/\allowbreak \%2e\%2e/\allowbreak gists}, and \texttt{DELETE /repos/\allowbreak acme/\allowbreak issues/1}), \emph{plus} two same-domain L7 outbound DLP exfiltration attacks on an allowlisted endpoint (\texttt{GET /repos/\allowbreak acme/\allowbreak issues?leak=AKIAIOSFODNN7EXAMPLE} and \texttt{POST /repos/\allowbreak acme/\allowbreak issues} with an OpenSSH PEM private key embedded in the JSON issue body). Unconfined, the raw token reaches the sink on all 10 requests (100.0\% ASR); in Aegis OS, per-secret domain binding, RFC 3986 path-canonicalized \texttt{SecretScopeRule} enforcement, and \texttt{inspect\_payload\_dlp} refused all 10 (0.0\% ASR).
    \item \textbf{Categories G and H (10 Conflicting Commits):} A commit batch holds a clean file (\texttt{a.txt}) and a file that a peer has changed on the same line (\texttt{b.txt}). Category G asks whether the clean half still reaches the host (broken atomicity); Category H asks whether the peer's edit is silently overwritten (lost update). Without transactions the unconfined baseline fails both; in Aegis OS Phase 1 refuses the whole batch, leaving the host tree untouched and the peer's edit intact (0.0\% ASR for both).
\end{itemize}

All 80 attacks were also run on Ubuntu 24.04 with identical results (0/80 for Aegis OS). By mounting an empty \texttt{--tmpfs} over every candidate host home directory (\texttt{\$HOME}, \texttt{/root}, \texttt{/home/*}, and \texttt{AOS\_EXTRA\_CREDENTIAL\_HOMES}) before selectively re-binding read-only compiler toolchains (\texttt{.rustup}, \texttt{.cargo/\{bin,registry,git\}}), Linux \texttt{bwrap} hides even unknown custom credential paths inside \texttt{\$HOME} (such as \texttt{\textasciitilde/custom\_vault/secret\_key.pem} in Category B). Note that the Linux sandbox requires unprivileged user namespaces, which Ubuntu 24.04 restricts by default through AppArmor (\texttt{kernel.apparmor\_restrict\_unprivileged\_userns}); we relaxed this in the test VM. Running the Linux benchmark also exposed two earlier defects that macOS runs could not: a private \texttt{/tmp} mounted after the writable binds hid every workspace under \texttt{/tmp}, which had made Linux categories A and B pass vacuously, and the benchmark counted a sandbox that merely failed to reach its proxy as compromised. Both are fixed; we report them because a security result that holds only when the sandbox is misconfigured is not a valid result.

\begin{table*}[!t]
\centering
\caption{\textbf{Experiment 2B (RQ2): Comparison with Stateless Native CLI Sandboxes} across all 80 attacks in Categories A--H, executed on both macOS (\texttt{sandbox-exec}) and Linux (\texttt{bwrap}) via \texttt{scripts/e5\_cli\_sandboxes.py} (identical counts on both OSes). \texttt{CLI-Sandbox (WS-Write)} mounts the live host repository read-write, carves out \texttt{.git} as read-only, and blocks \texttt{\textasciitilde/.ssh}, \texttt{\textasciitilde/.aws}, and \texttt{\textasciitilde/.gnupg}.}
\label{tab:clisandboxes}
\vspace{2pt}
\resizebox{\textwidth}{!}{%
\begin{tabular}{lcccccccccc}
\toprule
\textbf{Execution Mode} & \textbf{A: In-WS} & \textbf{B: Creds} & \textbf{C: Exfil} & \textbf{D: Tenant} & \textbf{E: Poison} & \textbf{F: Secret} & \textbf{G: Atomic} & \textbf{H: Overwrite} & \textbf{Total} & \textbf{ASR} \\
\midrule
\texttt{Host-YOLO} (Unconfined) & 10/10 & 10/10 & 10/10 & 10/10 & 10/10 & 10/10 & 10/10 & 10/10 & 80/80 & 100.0\% \\
\texttt{CLI-Sandbox (WS-Write + Net)} & 10/10 & 4/10 & 10/10 & 10/10 & 4/10 & 10/10 & 10/10 & 10/10 & 68/80 & 85.0\% \\
\texttt{CLI-Sandbox (WS-Write, No-Net)} & 10/10 & 4/10 & 0/10 & 10/10 & 4/10 & 0/10 & 10/10 & 10/10 & 48/80 & 60.0\% \\
\textbf{Aegis-OS (Transactional CoW)} & \textbf{0/10} & \textbf{0/10} & \textbf{0/10} & \textbf{0/10} & \textbf{0/10} & \textbf{0/10} & \textbf{0/10} & \textbf{0/10} & \textbf{0/80} & \textbf{0.0\%} \\
\bottomrule
\end{tabular}%
}
\end{table*}

\paragraph{Comparison with Stateless Native CLI Sandboxes (Experiment 2B).} Production agent CLIs (such as Codex CLI's \texttt{workspace-write} mode and Claude Code's bash sandbox) wrap shell commands in a stateless Seatbelt or Bubblewrap/Landlock profile that mounts the live working tree read-write, carves out \texttt{.git} as read-only, and blocks \texttt{\textasciitilde/.ssh}, \texttt{\textasciitilde/.aws}, and \texttt{\textasciitilde/.gnupg}. Table~\ref{tab:clisandboxes} evaluates this architecture on both macOS and Linux across Categories A--H. With outbound network enabled (\texttt{WS-Write + Net}, required for package managers and API calls), the stateless CLI sandbox still suffers an \textbf{85.0\% ASR (68/80)}: (i) in-workspace destructive commands (\texttt{rm -rf src}, \texttt{truncate}) directly corrupt the live host tree (\textbf{10/10} in Cat.~A); (ii) blocking \texttt{\textasciitilde/.ssh}, \texttt{\textasciitilde/.aws}, and \texttt{\textasciitilde/.gnupg} stops 6/10 credential reads, but unmasked dotfile credentials in \texttt{\$HOME} (\texttt{.npmrc}, \texttt{.pypirc}, \texttt{.netrc}, \texttt{.git-credentials}) remain readable (\textbf{4/10} in Cat.~B); (iii) raw \texttt{\$GITHUB\_TOKEN} environment variables are exfiltrated or sent to unbound domains (\textbf{10/10} in Cats.~C and F); (iv) carving out \texttt{.git}, \texttt{.envrc}, and \texttt{.github} as read-only blocks VCS/CI hooks (6/10), but in-tree build-system entrypoints (\texttt{build.rs}, \texttt{Makefile}, \texttt{package.json}, and \texttt{.cargo/config.toml}) remain writable in an in-place workspace sandbox (\textbf{4/10} in Cat.~E); and (v) direct in-place writes lack two-phase staging or 3-way merging (\textbf{10/10} in Cats.~G and H). Disabling outbound network entirely (\texttt{WS-Write, No-Net}) stops network exfiltration (Cats.~C and F) at the cost of breaking dependency installation and API access, yet still leaves \textbf{60.0\% ASR (48/80)} from in-place workspace mutations, dotfile reads, and concurrent overwrites.

\begin{table*}[!t]
\centering
\caption{\textbf{Experiment 3A (RQ3): Multi-Trial APFS Copy-on-Write Workspace Provisioning Latency and Storage Efficiency} across 10 concurrent agents modifying realistic per-task working sets (6, 12, and 25 files per agent; 5 trials per scale, Mean $\pm$ StdDev). The baseline is a user-space read/write loop, not a system copy tool; Table~\ref{tab:baselines} compares against real tools.}
\label{tab:exp3a}
\vspace{2pt}
\resizebox{\textwidth}{!}{%
\begin{tabular}{lcccccc}
\toprule
\textbf{Repository Scale} & \textbf{Files} & \textbf{Logical Size} & \textbf{User-Space Copy Loop (Mean $\pm$ SD)} & \textbf{Aegis CoW Clone (Mean $\pm$ SD)} & \textbf{Speedup vs.\ Loop} & \textbf{10-Agent Disk Saving} \\
\midrule
Small Project & 100 & 2.0 MB & $9.33 \pm 0.28$ ms & $\mathbf{0.71 \pm 0.05}$ \textbf{ms} & \textbf{13.2$\times$} & \textbf{94.0\%} (1.2 MB vs.\ 19.5 MB) \\
Medium Project & 300 & 17.6 MB & $32.08 \pm 5.82$ ms & $\mathbf{1.85 \pm 0.09}$ \textbf{ms} & \textbf{17.3$\times$} & \textbf{96.0\%} (7.0 MB vs.\ 175.8 MB) \\
Large Project & 1,000 & 78.1 MB & $113.13 \pm 16.41$ ms & $\mathbf{6.28 \pm 0.65}$ \textbf{ms} & \textbf{18.0$\times$} & \textbf{97.5\%} (19.5 MB vs.\ 781.2 MB) \\
\bottomrule
\end{tabular}%
}
\end{table*}

\begin{table}[!t]
\centering
\caption{\textbf{Experiment 3B (RQ3): Kernel, Daemon, Merge Commit, and Sandbox Spawn Latency Distribution}. Ops/sec is computed as $N$ divided by total wall-clock time of the measurement loop (including harness overhead), so it is slightly lower than $1/\text{Mean}$.}
\label{tab:exp3b}
\vspace{2pt}
\resizebox{\columnwidth}{!}{%
\begin{tabular}{lccccc}
\toprule
\textbf{Runtime Operation} & \textbf{Mean ($\mu$s)} & \textbf{P50 ($\mu$s)} & \textbf{P95 ($\mu$s)} & \textbf{P99 ($\mu$s)} & \textbf{Ops/sec} \\
\midrule
\texttt{PolicyKernel::evaluate} ($N{=}2000$) & 0.31 & 0.29 & 0.38 & 0.50 & 2,959,674 \\
\texttt{IpcBus::send + recv} ($N{=}1000$) & 1.06 & 0.83 & 1.79 & 4.75 & 924,749 \\
\texttt{HTTP POST /mcp} (\texttt{fs\_read}, $N{=}200$) & 177.11 & 183.08 & 258.79 & 409.00 & 5,503 \\
\texttt{commit\_to\_host} (3-Way Merge, $N{=}100$) & 424.02 & 413.08 & 540.21 & 646.12 & 2,110 \\
\texttt{Sandbox::exec} (\texttt{sandbox-exec}, $N{=}30$) & 6,678.97 & 6,557.62 & 7,533.12 & 7,811.33 & 150 \\
\bottomrule
\end{tabular}%
}
\end{table}

\subsection{Experiment 3 (RQ3): CoW Provisioning, Storage Footprint, and Operation Latency}

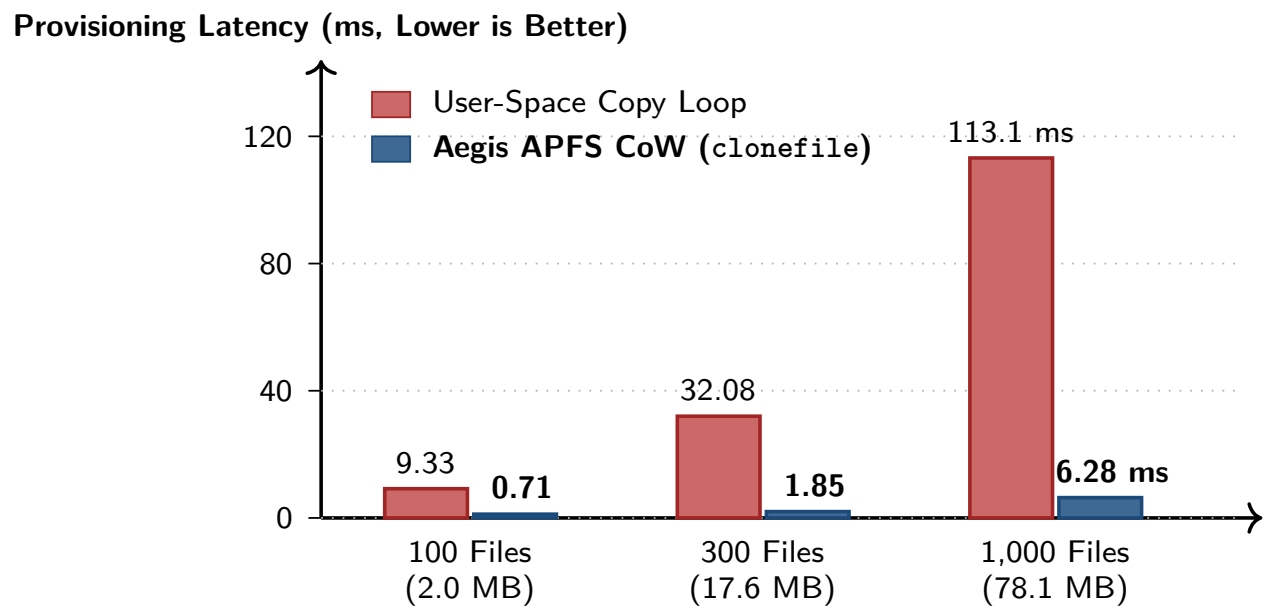
\begin{figure}[!t]
\centering
\resizebox{\columnwidth}{!}{%
\begin{tikzpicture}[font=\sffamily\scriptsize]
\draw[thick, ->] (0,0) -- (7.4,0) node[right] {};
\draw[thick, ->] (0,0) -- (0,3.6) node[above, align=center] {\textbf{Provisioning Latency (ms, Lower is Better)}};

\foreach \y/\label in {0/0, 1.0/40, 2.0/80, 3.0/120} {
    \draw (0,\y) -- (-0.1,\y) node[left] {\label};
    \draw[dotted, black!30] (0,\y) -- (7.2,\y);
}

\fill[aegisred!70, draw=aegisred!90!black, thick] (0.5,0) rectangle (1.15,0.23);
\node[above] at (0.82, 0.23) {9.33};
\fill[aegisblue!85, draw=aegisblue!95!black, thick] (1.2,0) rectangle (1.85,0.03);
\node[above] at (1.58, 0.03) {\textbf{0.71}};
\node[below, align=center] at (1.17, -0.05) {100 Files\\(2.0 MB)};

\fill[aegisred!70, draw=aegisred!90!black, thick] (2.8,0) rectangle (3.45,0.80);
\node[above] at (3.12, 0.80) {32.08};
\fill[aegisblue!85, draw=aegisblue!95!black, thick] (3.5,0) rectangle (4.15,0.05);
\node[above] at (3.88, 0.05) {\textbf{1.85}};
\node[below, align=center] at (3.47, -0.05) {300 Files\\(17.6 MB)};

\fill[aegisred!70, draw=aegisred!90!black, thick] (5.1,0) rectangle (5.75,2.83);
\node[above] at (5.42, 2.83) {113.1 ms};
\fill[aegisblue!85, draw=aegisblue!95!black, thick] (5.8,0) rectangle (6.45,0.16);
\node[above] at (6.22, 0.16) {\textbf{6.28 ms}};
\node[below, align=center] at (5.77, -0.05) {1,000 Files\\(78.1 MB)};

\fill[aegisred!70, draw=aegisred!90!black] (0.4, 3.15) rectangle (0.7, 3.35);
\node[right] at (0.75, 3.25) {User-Space Copy Loop};
\fill[aegisblue!85, draw=aegisblue!95!black] (0.4, 2.8) rectangle (0.7, 3.0);
\node[right] at (0.75, 2.9) {\textbf{Aegis APFS CoW (\texttt{clonefile})}};

\end{tikzpicture}%
}
\caption{Workspace provisioning latency across repository scales (Experiment 3A). APFS \texttt{clonefile} scales with file metadata count rather than total byte volume, achieving an 18.0$\times$ speedup over a user-space copy loop on the 1,000-file repository.}
\label{fig:exp3}
\end{figure}

We benchmark workspace provisioning across three repository scales (100 files / 2.0 MB, 300 files / 17.6 MB, and 1,000 files / 78.1 MB, 5 trials per scale). As shown in Table~\ref{tab:exp3a} and Figure~\ref{fig:exp3}, a user-space copy loop requires $113.13 \pm 16.41$ ms for the 1,000-file repository, whereas APFS \texttt{libc::clonefile} provisions the shadow workspace in \textbf{$6.28 \pm 0.65$ ms} (18.0$\times$ faster than that loop; Table~\ref{tab:baselines} gives the comparison against real tools). Because \texttt{clonefile} duplicates only inode metadata rather than file data blocks, provisioning latency scales with file count (0.71 ms at 100 files to 6.28 ms at 1,000 files) rather than total byte size. Across 10 concurrent agent workspaces modifying working sets of 6, 12, and 25 files per agent, APFS CoW reduces disk block allocation by \textbf{94.0\%--97.5\%} (19.5 MB vs.\ 781.2 MB at 1,000 files).

Table~\ref{tab:exp3b} reports the latency distribution across all five runtime operation classes. In-process policy evaluation takes \textbf{0.31 $\mu$s} (310 ns, 2.96M ops/s), P2P IPC send/receive takes \textbf{1.06 $\mu$s}, a loopback HTTP JSON-RPC \texttt{fs\_read} call (\texttt{POST /mcp}) takes \textbf{177.11 $\mu$s}, a two-phase LCS Diff3 \texttt{commit\_to\_host} takes \textbf{424.02 $\mu$s}, and spawning an external process under macOS Seatbelt (\texttt{/usr/bin/sandbox-exec /usr/bin/true}) takes \textbf{6.68 ms} (6,678.97 $\mu$s). Relative to typical LLM inference turns (0.5--2.0 s), MCP tool mediation adds $<0.04\%$ latency and Seatbelt process confinement adds $\sim$6.7 ms per spawned command.

\begin{table*}[!t]
\centering
\caption{\textbf{Experiment 3C (RQ3): Provisioning a 1,000-file, 78 MB Tree with Real Tools} (mean time in ms; 7 trials after a warm-up, methods interleaved; physical disk consumed in parentheses when above 1 MB). Plain copy tools receive the tree without \texttt{.git}, which Aegis OS also skips; the git-based methods inherently need the repository. External commands include 0.3--2.6 ms of process start-up; the Aegis figure is an in-process call (macOS: Table~\ref{tab:exp3a}; Linux: range over three runs of \texttt{aegis-os bench}). $^\dagger$XFS is a loop-backed image on btrfs and its Aegis timings are very noisy. ext4 has no reflink.}
\label{tab:baselines}
\vspace{2pt}
\resizebox{\textwidth}{!}{%
\begin{tabular}{lcccc}
\toprule
\textbf{Method} & \textbf{macOS APFS} & \textbf{Linux btrfs} & \textbf{Linux XFS$^\dagger$} & \textbf{Linux ext4} \\
\midrule
\texttt{cp -R} (default) & 205 (78 MB) & 9.9 & 11.5 & 30.6 (78 MB) \\
\texttt{cp -R --reflink=never} (true full copy) & = \texttt{cp -R} & 33.1 (78 MB) & 30.5 (79 MB) & 42.5 (78 MB) \\
\texttt{cp} with reflink (\texttt{-cR} / \texttt{--reflink=always}) & 99 & 9.7 & 10.6 & not supported \\
\texttt{rsync -a} & 335 (79 MB) & 82 (78 MB) & 75 (79 MB) & 93 (78 MB) \\
\texttt{git worktree add} & 125 (79 MB) & 57 (78 MB) & 62 (79 MB) & 64 (78 MB) \\
\texttt{git clone --local} & 382 (79 MB) & 69 (78 MB) & 71 (79 MB) & 107 (79 MB) \\
\midrule
\textbf{Aegis OS} (CoW clone) & \textbf{6.3} & \textbf{15.6} & 32.1 & 35.1 (78 MB) \\
\bottomrule
\end{tabular}%
}
\end{table*}

\paragraph{Comparison with Real Tools and Cross-OS Remote Daemon Evaluation (Experiment 3C).} The copy loop in Table~\ref{tab:exp3a} is a weak baseline, so Table~\ref{tab:baselines} times the tools a developer would actually use to give an agent an isolated tree. On APFS, Aegis OS (6.3 ms) is about 16$\times$ faster than \texttt{cp -cR} (99 ms), which clones file by file, and about 20$\times$ faster than \texttt{git worktree add} (125 ms); the advantage comes from cloning the whole directory in a single \texttt{clonefile} call. On Linux, GNU \texttt{cp -R} reflinks by default and takes 9.9 ms on btrfs against 15.6 ms for Aegis OS's single-open \texttt{FICLONE} traversal (3.2 ms vs.\ 4.4 ms at the 300-file medium scale), while on ext4 (no reflink) our single-open \texttt{copy\_file\_range} fallback provisions the 1,000-file tree in 35.1 ms (comparable to \texttt{cp -R} at 30.6 ms and faster than \texttt{cp --reflink=never} at 42.5 ms). Across all Linux filesystems, Aegis OS stays 2$\times$--5$\times$ faster than \texttt{rsync}, \texttt{git worktree add}, and \texttt{git clone --local}, while reflink-based methods allocate under 1 MB of extra disk. Finally, to verify a hybrid cloud/remote-container deployment (\texttt{scripts/verify\_remote\_linux\_daemon.py}), we ran \texttt{aegis-os daemon} inside the Ubuntu 24.04 Linux VM backed by \texttt{/mnt/btrfs} and drove a complete multi-agent workflow (\texttt{coder-alpha} and \texttt{reviewer-beta}) from the macOS host over TCP (\texttt{POST /v1/mcp}): provisioning CoW workspaces, minting scoped \texttt{ak\_live\_*} tenant keys, executing \texttt{proc\_exec} under \texttt{linux-bwrap-namespace-enforced} (verifying read-only root protection against \texttt{touch /etc/pwned}), approving HITL commit escalations via \texttt{POST /v1/approvals/decide}, and exporting a 3-way merged Git PR bundle (\texttt{POST /v1/workspaces/pr}). Across $N{=}50$ cross-OS \texttt{POST /v1/mcp} (\texttt{fs\_read}) round-trips from macOS to the Linux daemon, end-to-end latency averaged \textbf{0.80 ms} (P50 = \textbf{0.78 ms}, P99 = \textbf{0.99 ms}).

\begin{table*}[!t]
\centering
\caption{\textbf{Experiment 4 (RQ4): 3-Agent (\texttt{agy} + \texttt{claude} + \texttt{codex}) Concurrent Rust Modification, Syntactic \& Semantic Conflict Detection, and \texttt{cargo test} Verification}.}
\label{tab:exp4}
\vspace{2pt}
\resizebox{\textwidth}{!}{%
\begin{tabular}{llll}
\toprule
\textbf{Evaluation Metric} & \textbf{Host-Shared-YOLO (In-Place Write)} & \textbf{Aegis-OS Isolated CoW + 2-Phase LCS Diff3} & \textbf{Empirical Result} \\
\midrule
Concurrent 3-Agent Structural \& Function Edits (\texttt{src/lib.rs}) & 1 / 3 preserved (2 / 3 overwritten) & \textbf{3 / 3 preserved (100\% LCS Diff3 Merged)} & \textbf{3.0$\times$ Feature Retention (0\% Clobber)} \\
Overlapping Same-Line Edit Conflict Detection (10 Trials) & 0 / 10 detected (Silent Lost Update) & \textbf{10 / 10 detected \& aborted in Phase 1 (Host intact)} & \textbf{100\% Syntactic Conflict Safety} \\
Disjoint-Line Semantic Conflict Detection via \texttt{verify\_cmd} (10 Trials) & 0 / 10 detected (Broken Host Build) & \textbf{10 / 10 intercepted in Sandbox (Host 100\% Compilable)} & \textbf{100\% Semantic Build Integrity} \\
Post-Merge Real \texttt{cargo test --offline} Verification & Failed (Missing \texttt{agy} \& \texttt{claude} logic) & \textbf{Passed (\texttt{cargo test} exit code 0)} & \textbf{Build \& Test Verified} \\
In-Transit IPC Payload for 3$\times$64 KB Spec Handoffs & 196,608 bytes ($\sim$49,152 tokens inline) & \textbf{294 bytes ($\sim$73 tokens via CAS URI handles)} & \textbf{99.85\% In-Transit Token Reduction} \\
\bottomrule
\end{tabular}%
}
\end{table*}

\begin{table}[!t]
\centering
\caption{\textbf{Experiment 5 (RQ5): Multi-File Corruption Recovery and \texttt{cargo check} Verification} (5 trials/scale, Mean $\pm$ SD).}
\label{tab:exp5}
\vspace{2pt}
\resizebox{\columnwidth}{!}{%
\begin{tabular}{lcccc}
\toprule
\textbf{Corrupted Modules} & \textbf{Seq.\ File Restore} & \textbf{Aegis CoW Rollback} & \textbf{Speedup} & \textbf{\texttt{cargo check}} \\
\midrule
10 (640 KB) & $0.90 \pm 0.06$ ms & $\mathbf{0.75 \pm 0.04}$ \textbf{ms} & \textbf{1.2$\times$} & \texttt{PASSED} \\
50 (3.2 MB) & $4.82 \pm 0.32$ ms & $\mathbf{2.33 \pm 0.14}$ \textbf{ms} & \textbf{2.1$\times$} & \texttt{PASSED} \\
200 (12.8 MB) & $20.63 \pm 3.14$ ms & $\mathbf{8.68 \pm 1.02}$ \textbf{ms} & \textbf{2.4$\times$} & \texttt{PASSED} \\
\bottomrule
\end{tabular}%
}
\end{table}

\subsection{Experiment 4 (RQ4): Multi-Agent Concurrent Rust Coding and Conflict Detection}

We evaluate both disjoint structural edits and overlapping concurrent edits across \texttt{agy-agent}, \texttt{claude-code}, and \texttt{codex-cli} on a Rust crate (\texttt{src/lib.rs}):
\begin{itemize}
    \item \textbf{Concurrent Multi-Line Insertion and Function Edits (EXP 4A):} All three agents start from the same baseline \texttt{src/lib.rs}. \texttt{agy-agent} inserts a 3-line helper function (\texttt{min\_fee()}) at the top of the file and updates \texttt{calc\_fee} (shifting downstream line indices), while \texttt{claude-code} modifies \texttt{calc\_tax} and \texttt{codex-cli} modifies \texttt{calc\_discount}. When writing directly to a shared file (\textbf{Host-Shared-YOLO}), the final writer (\texttt{codex-cli}) overwrites the earlier edits, preserving only \textbf{1/3} of the features. In \textbf{Aegis-OS}, our two-phase LCS Diff3 engine aligns the unmodified anchor lines across the unequal-length files (matching the merge semantics of \texttt{git merge-file} without requiring manual \texttt{git worktree} branches per agent), preserving \textbf{3/3 (100\%)} of the features and passing \texttt{cargo test --offline} (\texttt{exit code 0}).
    \item \textbf{Overlapping Same-Line Conflict Detection (EXP 4B):} Across 10 trials, both \texttt{agy-agent} and \texttt{codex-cli} concurrently modify \texttt{calc\_fee} with conflicting expressions. While \textbf{Host-Shared-YOLO} suffers a silent lost update (0/10 detected), \textbf{Aegis-OS} detects \textbf{10/10 (100\%)} of the conflicts during Phase 1, aborts the second commit without modifying the host file, and returns the exact conflicting baseline line number.
    \item \textbf{Disjoint-Line Semantic Conflict Detection via Pre-Commit \texttt{verify\_cmd} Gate (EXP 4C):} Across 10 trials, \texttt{agy-agent} renames \texttt{calc\_fee} to \texttt{calc\_service\_fee} on line 1 while \texttt{codex-cli} concurrently adds a call to the old \texttt{calc\_fee} inside \texttt{calc\_tax} at the bottom of the file. Because the line ranges do not overlap, a purely syntactic Diff3 merge splices both changes together and produces an uncompilable file (0/10 detected in \textbf{Host-Shared-YOLO}). In \textbf{Aegis-OS}, passing \texttt{verify\_cmd = "rustc --crate-type lib --emit=metadata ..."} to \texttt{workspace\_commit} runs the compiler inside the agent's OS sandbox after \texttt{refresh\_from\_host()} merges the peer rename into the shadow workspace and \emph{before} any host file is touched, intercepting \textbf{10/10 (100\%)} of the semantic conflicts and leaving the host repository 100\% compilable until \texttt{codex-cli} updates its call site.
    \item \textbf{In-Transit IPC Token Savings:} Passing 98-byte \texttt{artifact://} CAS URI handles instead of inline 64 KB specifications across 3 inter-agent handoffs reduces in-transit IPC message payload bytes from 196,608 bytes ($\sim$49,152 tokens) to 294 bytes ($\sim$73 tokens), a \textbf{99.85\% reduction} for orchestrator routing and pass-through delegation (Table~\ref{tab:exp4}).
\end{itemize}

\subsection{Experiment 5 (RQ5): Time-Travel Rollback and \texttt{cargo check} Verification}

We evaluate recovery from multi-file syntax corruption (\texttt{BROKEN\_RUST\_SYNTAX\_ERROR !!!}) injected across 10, 50, and 200 Rust module files (\texttt{src/mod\_*.rs}, 64 KB per module). As shown in Table~\ref{tab:exp5}, Aegis OS's directory-tree \texttt{workspace\_rollback} restores all 200 corrupted modules in \textbf{$8.68 \pm 1.02$ ms} (\textbf{2.4$\times$ faster} than sequential file-by-file I/O restoration), and immediately passes \texttt{cargo check --offline} inside the restored shadow workspace across all scales.

\subsection{Experiment 6 (RQ6): Component Ablation}

To test whether each safety component is necessary, we build Aegis OS with a test-only \texttt{ablation} cargo feature that switches one component off through \texttt{AOS\_ABLATE}. In release builds the check is a compile-time \texttt{false}, so the switches cannot be set by an environment variable. We rerun all 80 attacks across all eight single-component ablations (Table~\ref{tab:ablation}); the macOS and Linux matrices are identical. Every individual component maps to the exact category it guards along a clean diagonal: without the OS kernel sandbox (\texttt{kernel\_sandbox}), path escape, credential harvesting, and direct socket bypass succeed 10/10; without credential-directory masking (\texttt{cred\_mask}), credential harvesting alone succeeds 10/10; without network/proxy socket confinement (\texttt{net\_isolation}), direct \texttt{--noproxy} egress bypass succeeds 10/10; without the commit-time guard (\texttt{commit\_guard}), 10/10 poisoned commits land under a lease; without the cross-tenant policy gate (\texttt{tenant\_gate}), 5/10 cross-tenant calls succeed; without per-secret domain/L7/DLP binding (\texttt{secret\_binding}), 10/10 tokens or payloads leak on allowlisted endpoints; without the three-way merge (\texttt{merge}), 10/10 peer edits are silently overwritten; and without Phase-1 pre-validation (\texttt{prevalidate}), 10/10 conflicting batches are half-applied.

Three observations qualify this result. First, the diagonal is partly by design, because each category was written to exercise one component; the table shows that each component is necessary for its category, not that the components do not overlap. Second, the one partial row (\texttt{tenant\_gate}: 5/10) is informative: disabling the policy tenant gate lets the five cross-tenant IPC sends through the policy layer, but none of the five CAS fetches succeed because the artifact store independently rejects foreign organization URIs. Motivated by this finding, we also added a broker-level tenant guard inside \texttt{AegisIpcBus::send} (\texttt{aos-ipc}) and restricted cross-repository workstation routing in \texttt{aos-mcp-gateway} to peers with verified live workstation OS PIDs on both endpoints. Third, ablating the commit-phase merge leaves Experiment 4A at 3/3, because \texttt{refresh\_from\_host} already merges non-overlapping peer edits before commit; the commit-phase merge specifically protects the concurrency window between refresh and commit, which Category H exercises.

\begin{table*}[!t]
\centering
\caption{\textbf{Experiment 6 (RQ6): Component Ablation.} Attacks (out of 10 per category) that compromise Aegis OS when a single component is switched off in a test-only build (\texttt{scripts/e5\_ablation.py}); identical on macOS and Linux. The first row is the full system.}
\label{tab:ablation}
\vspace{2pt}
\resizebox{\textwidth}{!}{%
\begin{tabular}{lcccccccc}
\toprule
\textbf{Component switched off (\texttt{AOS\_ABLATE})} & \textbf{A: Path} & \textbf{B: Creds} & \textbf{C: Egress} & \textbf{D: Tenant} & \textbf{E: Poison} & \textbf{F: Secret} & \textbf{G: Atomic} & \textbf{H: Overwrite} \\
\midrule
None (full system control) & 0 & 0 & 0 & 0 & 0 & 0 & 0 & 0 \\
OS kernel sandbox (\texttt{kernel\_sandbox}) & \textbf{10} & \textbf{10} & \textbf{10} & 0 & 0 & 0 & 0 & 0 \\
Credential-directory masking (\texttt{cred\_mask}) & 0 & \textbf{10} & 0 & 0 & 0 & 0 & 0 & 0 \\
Network / proxy socket confinement (\texttt{net\_isolation}) & 0 & 0 & \textbf{10} & 0 & 0 & 0 & 0 & 0 \\
Commit-time poisoning guard (\texttt{commit\_guard}) & 0 & 0 & 0 & 0 & \textbf{10} & 0 & 0 & 0 \\
Cross-tenant policy gate (\texttt{tenant\_gate}) & 0 & 0 & 0 & \textbf{5} & 0 & 0 & 0 & 0 \\
Per-secret L7 scope \& outbound DLP (\texttt{secret\_binding}) & 0 & 0 & 0 & 0 & 0 & \textbf{10} & 0 & 0 \\
Three-way merge (\texttt{merge}) & 0 & 0 & 0 & 0 & 0 & 0 & 0 & \textbf{10} \\
Phase-1 pre-validation (\texttt{prevalidate}) & 0 & 0 & 0 & 0 & 0 & 0 & \textbf{10} & 0 \\
\bottomrule
\end{tabular}%
}
\end{table*}

\begin{table*}[!t]
\centering
\caption{\textbf{Cross-Platform Operating System Primitive Mapping and Implementation Status} in Aegis OS.}
\label{tab:crossplatform}
\vspace{2pt}
\resizebox{\textwidth}{!}{%
\begin{tabular}{lllll}
\toprule
\textbf{Aegis OS Subsystem} & \textbf{macOS Workstation (Evaluated)} & \textbf{Linux Workstation (Evaluated, VM)} & \textbf{Cloud Host (Tier 2)} & \textbf{Current Implementation Status} \\
\midrule
\textbf{CoW Workspace (\texttt{aos-vfs})} & APFS \texttt{libc::clonefile(src, dst, 0)} on whole directories & Per-file \texttt{FICLONE} \& optional \texttt{BTRFS\_IOC\_SNAP\_CREATE\_V2}; \texttt{copy\_file\_range} (ext4) & Virtio-fs CoW mount spec & Both evaluated (Table~\ref{tab:baselines}) \\
\textbf{Process MAC (\texttt{aos-sandbox})} & XNU Seatbelt (\texttt{/usr/bin/sandbox-exec}) & Bubblewrap (\texttt{/usr/bin/bwrap --unshare-all}, private \texttt{/tmp}) & KVM (\texttt{libkrun} / Firecracker) & Seatbelt and \texttt{bwrap} evaluated; Tier-2 spec \\
\textbf{Credential Blocking (\texttt{\textasciitilde/.ssh})} & Seatbelt \texttt{(deny file-read* (regex ...))}, any path & Default-deny \texttt{--tmpfs} over \texttt{\$HOME}/\texttt{/root}/\texttt{/home/*} + safe toolchain re-binds & Unmounted host volumes & Both evaluated (blocks unknown \texttt{\$HOME} paths) \\
\textbf{IPC / GUI Escape Guard} & Seatbelt \texttt{(deny appleevent-send)} & \texttt{bwrap} PID/IPC namespace isolation & Guest VM kernel boundary & Implemented (Landlock/\texttt{seccomp-bpf} planned) \\
\textbf{Zero-Trust Egress (\texttt{aos-egress})} & Pinned TCP Proxy + Seatbelt \texttt{localhost:<port>} + L7 DLP & Pinned TCP Proxy + netns + token Unix-socket bridge + L7 DLP & Tap/vsock Egress Gateway & Evaluated on macOS, Linux \& Cross-OS TCP \\
\bottomrule
\end{tabular}%
}
\end{table*}

\section{Discussion and Limitations}
\label{sec:discussion}

\paragraph{Cross-Platform Implementation Scope (macOS vs.\ Linux).} Table~\ref{tab:crossplatform} summarizes each OS primitive. Both platforms were evaluated end-to-end (\S\ref{sec:eval}), Linux inside a virtual machine as well as in a cross-OS macOS-client-to-Linux-daemon configuration (\texttt{scripts/verify\_remote\_linux\_daemon.py}). While Seatbelt denies credential reads by regex path pattern and confines outbound sockets to one loopback port, our Linux \texttt{bwrap} driver achieves equivalent \texttt{\$HOME} credential isolation by mounting an empty \texttt{--tmpfs} over every candidate host home directory (\texttt{\$HOME}, \texttt{/root}, \texttt{/home/*}, and \texttt{AOS\_EXTRA\_CREDENTIAL\_HOMES}) before selectively re-binding read-only compiler toolchains (\texttt{.rustup} and \texttt{.cargo/\{bin,registry,git\}}), hiding both known and unknown credential paths inside user homes (verified on \texttt{\textasciitilde/custom\_vault/secret\_key.pem}); credentials placed in world-readable system directories outside user homes and \texttt{/tmp} still require listing in \texttt{AOS\_EXTRA\_CREDENTIAL\_HOMES}. Linux default cloning uses our 7-syscall \texttt{FICLONE} traversal (15.6 ms on btrfs) or optional \texttt{BTRFS\_IOC\_SNAP\_CREATE\_V2} subvolume snapshots, and falls back to \texttt{copy\_file\_range} (35.1 ms) on filesystems without reflink (ext4). A Linux sandbox additionally requires unprivileged user namespaces, which Ubuntu 24.04 disables for unprivileged programs by default through AppArmor; a deployment must either relax \texttt{kernel.apparmor\_restrict\_unprivileged\_userns} or ship an AppArmor profile for bwrap. Landlock, \texttt{seccomp-bpf}, and binding \texttt{MicroVmSandboxDriver} to live \texttt{libkrun}/Firecracker processes remain future work.

\paragraph{Scope and Validity of the Evaluation.} Several considerations bound what the numbers show. (i) While Experiment 1 replays 50 deterministic trajectory templates and Experiment 1B evaluates 52 real coding-agent session transcripts totaling 28,087 tool calls (plus 5 live Aegis OS dogfooding sessions), our real-trace corpus comes from a single engineering organization; broader field studies across diverse developer populations remain valuable future work. (ii) The 80 attacks in Experiment 2 were constructed to test the eight threat categories. The \texttt{Host-YOLO} column is 100\% by construction and the ablation diagonal is partly by design (Experiment 6); this evidence shows that each mechanism works as specified and is non-redundant for its category, not that no unmodeled kernel or toolchain bypass exists. (iii) The macOS latencies come from one Apple Silicon workstation, the Linux numbers from a virtual machine on the same host, and the XFS volume was a loop-backed image.

\paragraph{macOS Seatbelt Architecture and Escape Surface.} On macOS, dynamic linker caches (\texttt{dyld}), Xcode toolchains, and Homebrew packages traverse numerous paths under \texttt{/System}, \texttt{/Library}, and \texttt{/opt/homebrew}. Thus, our Seatbelt profile pairs a strict write allowlist (\texttt{(deny file-write*)}) with a regex read denylist for credential stores, \texttt{(deny appleevent-send)} to prevent AppleScript/GUI automation escapes, and exact port confinement \texttt{(allow network-outbound (remote ip "localhost:<proxy\_port>"))}. While Apple has marked \texttt{/usr/bin/sandbox-exec} as a legacy interface, it remains the standard kernel MAC mechanism used by production agent CLIs (including Claude Code and Codex CLI) on macOS because Apple's Endpoint Security and \texttt{Virtualization.framework} require root entitlements or full VM boot overhead. Hardening Mach bootstrap lookups (\texttt{mach-lookup}) against LaunchServices job spawning without breaking toolchain execution is an important direction for future work.

\paragraph{Session Leases and Commit-Time Poisoning.} Granting a 1-hour session lease (\texttt{ttl=3600}) eliminates 99.88\% of prompts on controlled trajectories (and 97.9\% on real session traces) by allowing validated CoW diffs to commit automatically. To prevent a prompt-injected agent from silently modifying repository hooks or build scripts under an active lease (\emph{commit-time poisoning}), Rule R0b (\S\ref{sec:policy}) refuses silent lease auto-commits whenever a diff touches VCS/shell hooks (\texttt{.git/hooks/*}, \texttt{.git/config}, \texttt{.gitmodules}, \texttt{.envrc}), CI pipelines (\texttt{.github/workflows/*}, \texttt{.gitlab-ci.yml}), or build-system and IDE entrypoints (\texttt{build.rs}, \texttt{Makefile}, \texttt{GNUmakefile}, \texttt{CMakeLists.txt}, \texttt{package.json}, \texttt{setup.py}, \texttt{.cargo/config*}, \texttt{.vscode/tasks.json}). A remaining open vector is indirect poisoning of application source files that are subsequently executed on the host outside Aegis OS without human diff review; running host builds inside \texttt{proc\_exec} or reviewing committed diffs mitigates this residual risk.

\paragraph{Confused Deputy on Allowlisted Egress Domains and HTTPS TLS Termination.} Boundary secret injection guarantees that raw tokens never enter the agent sandbox and cannot be sent to unauthorized domains. Furthermore, to defeat same-domain confused-deputy attacks on multi-tenant API hosts (such as an agent using a token bound to \texttt{api.github.com} to exfiltrate code via \texttt{POST /gists}, path traversal \texttt{/repos/acme/../../gists}, or embedding a key in an issue body), \texttt{aos-egress} enforces RFC 3986 path canonicalization (\texttt{normalize\_request\_path}), per-secret L7 HTTP method and URL path-prefix capability rules (\texttt{SecretScopeRule}), and an outbound DLP payload scanner (\texttt{inspect\_payload\_dlp}) that inspects URL query strings and HTTP bodies (up to 64 KiB) for secret placeholders, registered vault values, and standard credential formats (PEM private keys, AWS \texttt{AKIA*}, GitHub \texttt{ghp\_*}/\texttt{github\_pat\_*}, \texttt{sk-*}, and \texttt{xoxb-*} tokens, evaluated in Category F). Two residual limitations remain: (i) if an agent is legitimately granted write access to an endpoint (e.g., \texttt{POST /repos/acme/app/issues}), a prompt-injected payload could still exfiltrate arbitrary proprietary source code or obfuscated/base64-encoded strings that do not match structured secret signatures; and (ii) intercepting headers and bodies inside direct \texttt{HTTPS CONNECT} tunnels from legacy CLI binaries requires configuring the local proxy with a session-scoped ephemeral TLS Certificate Authority (\texttt{SSL\_CERT\_FILE}).

\paragraph{Syntactic 3-Way Merging vs.\ Pre-Commit Semantic Verification.} Our two-phase LCS Diff3 engine resolves disjoint line edits and aborts on overlapping line conflicts, while the optional sandboxed pre-commit verification gate (\texttt{verify\_cmd} in \texttt{workspace\_commit}, evaluated in Experiment 4C) catches disjoint-line semantic conflicts~\cite{apel2011semistructured} inside the agent's OS sandbox before any host file is modified. The trade-off is verification latency: running fast type/metadata checks (\texttt{rustc --emit=metadata} or incremental \texttt{cargo check}) adds tens to hundreds of milliseconds at commit time, whereas full end-to-end integration test suites on large monorepos may take seconds to minutes.

\section{Related Work}
\label{sec:related}

\paragraph{Operating System Transactions and Speculative Execution.} Our transaction-boundary security model builds on classical operating system transactions. TxOS~\cite{porter2009txos} introduced kernel-level system transactions (\texttt{sys\_xbegin}/\texttt{sys\_xend}) providing ACID semantics for OS resources, while Speculator~\cite{nightingale2005speculator} enabled speculative process execution over distributed filesystems with rollback on dependency failure. Mbox~\cite{kim2013mbox} overlays a private sandbox filesystem on the host via system-call interposition so that non-root users can inspect and selectively commit changes, and UnionFS~\cite{wright2004unionfs} unifies multiple directories into a single copy-on-write namespace inside the kernel. Aegis OS adapts OS transaction semantics specifically to multi-tenant AI coding agents, combining APFS \texttt{clonefile} CoW shadow trees with two-phase LCS Diff3 merging~\cite{khanna2007formal,apel2011semistructured} and Cedar-inspired~\cite{cutler2024cedar} session leases.

\paragraph{Authorization Fatigue in Usable Security.} Empirical studies of mobile permission dialogs~\cite{felt2012android} and browser security warnings~\cite{akhawe2013alice} show that users frequently ignore, misunderstand, or click through security prompts, and that prompt design and frequency strongly affect whether warnings are heeded. Autonomous coding agents exacerbate this phenomenon by issuing dozens of mutating tool calls per minute; Aegis OS eliminates intermediate prompts by making in-sandbox actions side-effect-free and reversible.

\paragraph{LLM Agent Sandboxes and Prompt Injection Defenses.} Benchmarks such as SWE-bench~\cite{jimenez2024swebench}, SWE-agent~\cite{yang2024sweagent}, OpenHands~\cite{wang2025openhands}, ToolEmu~\cite{ruan2024toolemu}, AgentDojo~\cite{debenedetti2024agentdojo}, and InjecAgent~\cite{zhan2024injecagent} demonstrate both the utility of coding agents and their susceptibility to indirect prompt injection~\cite{greshake2023not}. System-level defenses such as IsolateGPT~\cite{wu2024secgpt}, CaMeL~\cite{debenedetti2025camel}, and FIDES~\cite{costa2025fides} enforce privilege separation or information-flow control across LLM planners and tool interpreters. Production CLIs such as Claude Code (\texttt{sandbox-runtime}) and Codex CLI integrate Seatbelt and Bubblewrap/Landlock sandboxes to confine shell commands in place. Aegis OS complements both model-level IFC defenses and in-place CLI sandboxes by adding transactional CoW shadow workspaces, two-phase Diff3 multi-agent merge commits, boundary secret injection, and cross-agent IPC/CAS.

\paragraph{Agent Operating Systems and Multi-Agent Frameworks.} AIOS~\cite{mei2024aios} schedules LLM queries and context snapshots above application agents, while AutoGen~\cite{wu2023autogen} and MetaGPT~\cite{hong2024metagpt} coordinate multi-agent dialogues in Python. Aegis OS instead operates purely as a language-agnostic OS runtime bridging Anthropic's Model Context Protocol (MCP)~\cite{anthropic2024mcp} for vertical tool mediation and Google's Agent2Agent (A2A) protocol~\cite{google2025a2a} for horizontal capability discovery and stateful task lifecycles, while drawing on capability systems such as Capsicum~\cite{watson2010capsicum} and microVMs such as Firecracker~\cite{agache2020firecracker}.

\section{Conclusion}
\label{sec:conclusion}

We presented \textbf{Aegis OS}, a transactional, multi-tenant Agent Operating System Runtime that addresses the Security--Autonomy Dilemma for autonomous coding agents. By combining Copy-on-Write shadow workspaces, two-phase LCS Diff3 multi-agent merge commits with sandboxed pre-commit semantic verification (\texttt{verify\_cmd}), kernel-enforced Seatbelt/Bubblewrap sandboxes with default-deny \texttt{--tmpfs \$HOME} credential isolation and port-pinned network confinement, L7 method/path-scoped boundary secret injection with outbound DLP payload inspection, and intra-organization A2A IPC + CAS primitives, Aegis OS reduces human confirmation prompts by 94.19\%--99.88\% on controlled trajectories and by \textbf{90.7\%--97.9\% across 28,087 tool calls in 52 real coding-agent sessions}, while blocking all 80 evaluated attacks across eight categories on macOS and Linux. A nine-row single-component ablation confirms that each safety mechanism is necessary for the attack class it guards.

\bibliographystyle{plain}
\bibliography{references}

\appendix

\section{Appendix: Reproducibility and Artifact Guide}
\label{sec:appendix}

The open-source implementation of Aegis OS and the benchmark suite are available at \url{https://github.com/heurion-org/aegis-os}. All nine Rust crates (\texttt{aos-core}, \texttt{aos-policy}, \texttt{aos-vfs}, \texttt{aos-ipc}, \texttt{aos-sandbox}, \texttt{aos-egress}, \texttt{aos-recorder}, \texttt{aos-mcp-gateway}, and \texttt{aos-cli}) are self-contained in the workspace. To run all unit tests and reproduce all empirical experiments (\texttt{BENCHMARK\_REPORT.md} and \texttt{BENCHMARK\_REPORT.json}):
\begin{lstlisting}
cargo test --workspace
cargo run --release --bin aegis-os -- bench \
  --report-out BENCHMARK_REPORT.md
\end{lstlisting}
The real-trace evaluation (Experiment 1B), stateless CLI sandbox comparison (Experiment 2B), real-tool provisioning baselines (Experiment 3C), and component ablation matrix (Experiment 6) are reproduced via self-contained scripts in \texttt{scripts/}:
\begin{lstlisting}
python3 scripts/e6_real_traces.py
python3 scripts/e5_cli_sandboxes.py
python3 scripts/e5_baselines.py --base <dir on the filesystem under test>
cargo build --release -p aos-cli --features ablation
python3 scripts/e5_ablation.py --bin target/release/aegis-os
python3 scripts/verify_remote_linux_daemon.py
\end{lstlisting}

\typeout{get arXiv to do 4 passes: Label(s) may have changed. Rerun}
\end{document}